\documentclass[iop, apj, twocolappendix, numberedappendix, appendixfloats]{emulateapj}
\usepackage{threeparttable}
\usepackage{multirow}
\usepackage{times}
\usepackage{enumitem}
\usepackage{url} 
\usepackage{amsmath,bm}

\usepackage[T1]{fontenc}

\usepackage{upgreek}
\usepackage{threeparttable}
\usepackage{amsmath}
\usepackage{verbatim}

\shorttitle{The Circular Velocity Curve of the Milky Way using LRGB stars}
\shortauthors{Zhou et al.}

\begin{document}

\title{The Circular Velocity Curve of the Milky Way from 5 to 25\,kpc using luminous red giant branch stars}

\author{Yuan Zhou\altaffilmark{1,6}}
\author{Xinyi Li\altaffilmark{1,6}}
\author{Yang Huang\altaffilmark{2,3,7}}
\author{Huawei Zhang\altaffilmark{4,5}}

\altaffiltext{1}{South-Western Institute for Astronomy Research, Yunnan University, Kunming 650500, People's Republic of China; bashiyi@163.com; li\_xinyi621@mail.ynu.edu.cn}
\altaffiltext{2}{University of Chinese Academy of Sciences, Beijing 100049,  People's Republic of China; huangyang@bao.ac.cn}
\altaffiltext{3}{National Astronomical Observatories, Chinese Academy of Sciences, Beijing 100012, People’s Republic of China;}
\altaffiltext{4}{Department of Astronomy, Peking University, Beijing 100871, People’s Republic of China}
\altaffiltext{5}{Kavli Institute for Astronomy and Astrophysics, Peking University, Bejing 100871, People’s Republic of China}
\altaffiltext{6}{Co-first author: These authors contributed equally to this work.}
\altaffiltext{7}{Corresponding authour: Yang Huang}

\begin{abstract}
We present a sample of 254,882 luminous red giant branch (LRGB) stars selected from the APOGEE and LAMOST surveys. By combining photometric and astrometric information from the 2MASS and Gaia surveys, the precise distances of the sample stars are determined by a supervised machine learning algorithm: the gradient boosted decision trees. To test the accuracy of the derived distances, member stars of globular clusters (GCs) and open clusters (OCs) are used. The tests by cluster member stars show a precision of about 10 per cent with negligible zero-point offsets, for the derived distances of our sample stars. The final sample covers a large volume of the Galactic disk(s) and halo of $0<R<30$\,kpc and $|Z|\leqslant15$\,kpc. The rotation curve (RC) of the Milky Way across radius of $5\lesssim R\lesssim25$\,kpc have been accurately measured with $\sim$\,54,000 stars of the thin disk population selected from the LRGB sample. The derived RC shows a weak decline along $R$ with a gradient of $-1.83\pm0.02$\,$({\rm stat.}) \pm 0.07$\,$({\rm sys.})$\,km\,s$^{-1}$\,kpc$^{-1}$, in excellent agreement with the results measured by previous studies. The circular velocity at the solar position, yielded by our RC, is $234.04\pm0.08$\,$({\rm stat.}) \pm 1.36$\,$({\rm sys.})$\,km\,s$^{-1}$, again in great consistent with other independent determinations. From the newly constructed RC, as well as constraints from other data, we have constructed a mass model for our Galaxy, yielding a mass of the dark matter halo of $M_{\rm{200}}$ = ($8.05\pm1.15$)$\times$10$^{11} \rm{M_\odot}$ with a corresponding radius of $R_{\rm{200}}$ = $192.37\pm9.24$\,kpc and a local dark matter density of $0.39\pm0.03$\,GeV\,cm$^{-3}$.
\end{abstract}


\section{Introduction} 
The rotation curve (RC) represents the circular speed at different radial distance $R$ from the center of a galaxy, and therefore the RC measurement provides strong constraints on the galactic mass distribution, especially the dark content. Pioneered by \citet{1980ApJ...238..471R}, flat rotation curves (RCs) for spiral galaxies show a clear discrepancy between the observed velocities and what expected from the known mass distribution. This discrepancy is generally explained as due to the presence of dark matter in galaxies. For the Milky Way, the Galactic RC is a widely used tool to infer the mass distribution of different structural components, and provides vital clues to understand various fundamental issues, such as the local dark matter density \citep[]{2010A&A...523A..83S,2010A&A...509A..25W,2012PASJ...64...75S,2015MNRAS.453..377W}. 

However, it is quite challenging to accurately measure the RC of our Galaxy \citep[e.g.][]{2015MNRAS.453..377W,2020SCPMA..6309801W}. For the inner region inside the solar radius, the profile of Galactic RC has been determined by the tangent-point (TP) approach, typically based on the radio observations of gas emissions from the H{\sc i} 21\,cm or the CO 2.6\,mm \citep[]{1978A&A....63....7B,1985ApJ...295..422C,1989ApJ...342..272F,2008ApJ...679.1288L,2009PASJ...61..227S}. This method is under the assumption that gas moves around the Galactic center in purely circular orbits. The inferred RC by the TP method is therefore not reliable in the most central region, i.e., $R < $4-5\,kpc, due to the significant perturbations on the orbits of gas from the central bar \citep{2015A&A...578A..14C}. For the outer region beyond the solar radius, the Galactic RC has been determined by various tracers with kinematic information (radial velocity or proper motions or both) and known distances, e.g., the standard candles of classical cepheids \citep{1997A&A...318..416P} and red clump giants \citep[]{2012ApJ...759..131B,2014A&A...563A.128L,2016MNRAS.463.2623H}, the H{\sc ii} regions \citep[]{1989ApJ...342..272F,1993A&A...275...67B}, the planetary nebulae \citep{1983ApJ...274L..61S}, the masers \citep[]{2012PASJ...64..136H,2014ApJ...783..130R}, the blue horizontal branch stars \citep[]{2008ApJ...684.1143X,2012MNRAS.424L..44D,2012ApJ...761...98K} and the globular clusters or dwarf galaxies \citep[]{2009MNRAS.392L...1S,2010MNRAS.406..264W,2013ApJ...768..140B,2017ApJ...850..116L,2020ApJ...894...10L}. However, due to the lack of accurate proper motions and limited precisions of distances, the constructed RC profiles have suffered both large random and systematic errors \citep[e.g.][]{2015MNRAS.453..377W,2020SCPMA..6309801W}.

Most recently, this situation has been significantly improved, thanks to the release of Gaia data, which provides accurate astrometric information, i.e., parallax and proper motions, for over one billion stars in our Galaxy \citep[]{2016A&A...595A...2G,2018A&A...616A...1G}. With the help of accurate measurements of parallax from the Gaia DR2 \citep{2018A&A...616A...2L}, \citet[][hereafter H19]{2019AJ....158..147H} present a sample of $\sim$\,40,000 luminous red giant branch (LRGB) stars, selected from the APOGEE DR\,14 \citep{2017AJ....154...94M}, with distances as far as 20\,kpc from the Sun and a claimed precision of 10\%. \citet[][hereafter E19]{2019ApJ...871..120E} have measured the RC of our Galaxy from 5\,kpc to 25\,kpc based on a subsample of $\sim$\,23,000 thin disk stars taken from the LRGB sample of H19. Although the RC has been significantly improved by E19, the uncertainties (both systematic and random) are still quite large for the outer part of the profile (see Fig.\,3 of E19) that is vital to constrain the mass distribution of our Galaxy, in particular the dark matter component. 

Inspired by H19 and E19, we present the largest sample of nearly 260,000 LRGB stars (6-7 times larger than the sample by H19), selected from the APOGEE and LAMOST spectroscopic surveys, with precise distance measurements owing to the accurate parallax measurements from the Gaia EDR3 as training. By $\sim$\,54,000 thin disk stars (2-3 times larger than the adopted sample by E19) selected from our LRGB sample, a new accurate RC curve is constructed across the Galactocentric radius of 5 $\lesssim$ $R$ $\lesssim$ 25\,kpc, with significant improvements for the outer part. The paper is organized as follows. We briefly describe our data and samples in Section\,2. The determinations of distances of our sample stars are described in Section\,3. In Section\,4, the construction of the Galactic RC is described in detail. Given the constraints from the newly constructed RC and other data, a mass model of various components is built for our Galaxy in Section\,5. Finally, a summary is presented in Section\,6. 

\section{data}
\subsection{Coordinate systems}
In this paper, two coordinate systems are adopted. We use a right-handed Cartesian Galactocentric coordinate system $(X, Y, Z)$, with $X$ pointing towards the Galactic center, $Y$ in the direction of Galactic rotation and $Z$ towards the north Galactic pole. In addition, we adopt the standard Galactocentric cylindrical system ($R$, $\phi$, $Z$) with $R$ increasing radially outward, $\phi$ towards the Galactic rotation direction and $Z$ the same as that in the Cartesian system. We adopt the distance from the Sun to the Galactic center $R_{0} = 8.122 \pm 0.031$\,kpc \citep{2018A&A...615L..15G}, the vertical distance to the Galactic plane $z_\odot$ = 25\,pc \citep{2008ApJ...673..864J}, and the Galactocentric velocity of the Sun $V_{R,\odot} =$ 11.1\,km\,s$^{-1}$ \citep{2010MNRAS.403.1829S}, $V_{\phi,\odot} = $ 245.6\,km\,s$^{-1}$ and $V_{Z,\odot} =$ 7.8\,km\,s$^{-1}$, yielded by the proper motion of the sgr A$^*$ \citep{2004ApJ...616..872R} and the adopted value of $R_{0}$.

\subsection{Data}

\subsubsection{Spectroscopic surveys}
This study makes use of the spectra and stellar parameters from the APOGEE \citep{2017AJ....154...94M} and LAMOST surveys \citep[]{2012RAA....12..735D,2012RAA....12.1243L}.

As an important component of the SDSS III and IV \citep[]{2011AJ....142...72E,2017AJ....154...28B}, the APOGEE survey collects near-infrared high-resolution ($R \sim$\,22,500) spectra with wavelength covering the $H$-band (1.51-1.70\,$\upmu$m), mainly targeting the red giant stars in the Galactic disk and bulge, by two identical spectrographs installed in two 2.5\,m telescopes at the Apache Point Observatory and the Las Campanas Observatory, respectively. Here, we adopt the data from the APOGEE DR17 \citep{2022ApJS..259...35A}, the latest and final one, that includes stellar atmospheric parameters (i.e., effective temperature $T_{\rm eff}$, surface gravity log\,$g$ and metallicity [Fe/H]) and over 20 element abundance ratios for about 734,000 stars. 

The LAMOST survey collects both low ($R \sim$\,2000) and medium resolution spectra ($R \sim$\,7500) with a 4-metre quasi-meridian reflecting Schmidt telescope equipped with 4000 fibers distributed in a field of view of 5 degree in diameter \citep{2012RAA....12.1197C}. In the present study, we adopt the low-resolution spectra data from the LAMOST DR8\footnote{\url{http://www.lamost.org/dr8/}}. In this release, over ten million low-resolution optical (3000-9000\,\AA) spectra and the stellar atmospheric parameters derived from those spectra are available. In addition to the atmospheric parameters provided by the official release,  we also adopt the values of [$\alpha$/Fe] determined from the low-resolution by \citet{2022AJ....163..149W}.

\subsubsection{Astrometric and photometric surveys}
In addition to the aforementioned spectroscopic surveys, the data from the Gaia Early Data Release 3 \citep[][EDR3]{2021A&A...649A...1G} is also used in this study. Gaia EDR3 has published accurate astrometric and photometric data for over 1.8 billion astrophysical sources, with $G$-band magnitude from 3 to 21. For our targets, the typical uncertainties of parallax and proper motion measurements in Gaia EDR3 are 0.02-0.03\,mas and 0.02-0.03\,mas\,yr$^{-1}$, respectively.

To derive the distances of our LRGB sample stars, $J$ and $K_{\rm s}$ bands photometric data from the 2MASS survey \citep{2006AJ....131.1163S} are also used. The apparent magnitudes $J$ and $K_{\rm s}$ have been corrected for the interstellar extinction. The adopted values of extinction are estimated using the methods detailed described in Appendix A. The extinction coefficients for $J$ and $K_{\rm s}$ bands are taken from \citet{2013MNRAS.430.2188Y}.

\subsection{LRGB samples}
In this study, the LRGB stars are chosen as our tracers to derive the Galactic RC. As their top locations on the Herzsprung-Russell (HR) diagram, the LRGBs are more luminous than the usual disk tracers: red clump stars. Meanwhile, the LRGBs are very abundant with almost a full coverage of stellar populations. As discussed by H19, the LRGBs can be thought as standardizable candles since their luminosities can be predicted by simple functions of their stellar properties, including chemical compositions, effective temperature, surface gravity and stellar age. Taking advantage of those large-scale Galactic surveys, the required properties (to predict the luminosities/absolute magnitudes) of LRGBs can be measured from the stellar spectra and $K_{\rm s}$ band photometric data (see Section\,3 for details). The LRGBs are empirically on top of the red clump stars, that are a significant clustering feature on the HR diagram (see Fig.\,1). As the same as H19, we select the LRGBs from the APOGEE DR\,17 using the following cut:
\begin{equation}
0 \leqslant {\rm log}\,g \leqslant 2.2.
\end{equation}
For the LAMOST data, the feature of red clump stars on plane of $T_{\rm eff}$-$\rm{log}$\,$g$ is different from that of the APOGEE data, which is largely due to the systematic differences of the atmospheric parameters delivered by the two surveys. We thus empirically develop a new cut to select LRGBs from the LAMOST data:
\begin{equation}
0 \leqslant {\rm log}\,g \leqslant 0.002\,T_{\rm{eff}}- 8.138.
\end{equation}
The cuts to select LRGBs from the APOGEE and LAMOST are shown in Fig.\,1. In total, 154,534 and 157,312 LRGB stars with spectral signal-to-noise ratios (SNRs) greater than 20, together with photometric (i.e., $J$ and $K_{\rm s}$ bands) and astrometric (i.e., parallax and proper motions) information, are obtained.

\begin{figure*}
\begin{center}
\includegraphics[width=8.8cm,height=8.5cm]{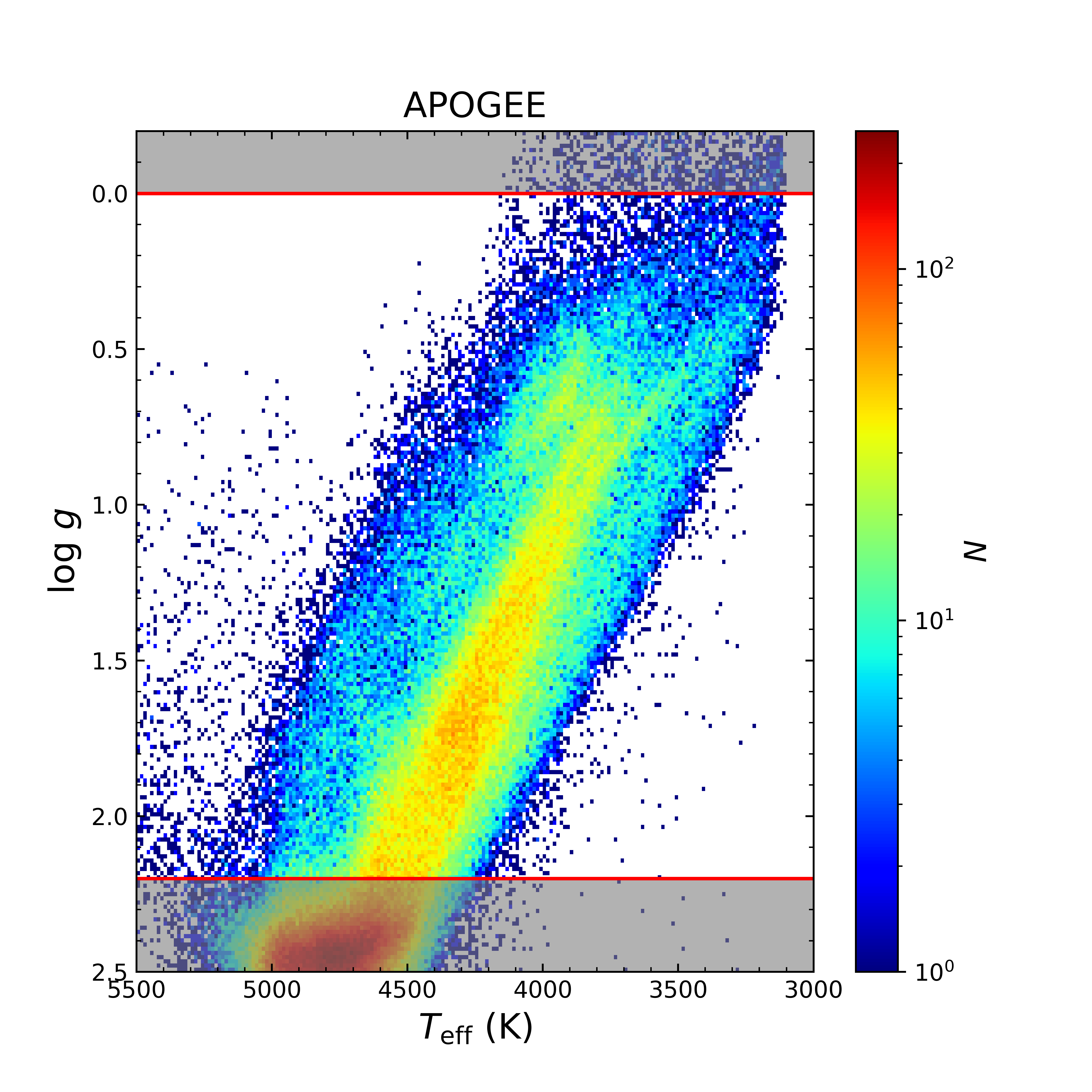}
\includegraphics[width=8.8cm,height=8.5cm]{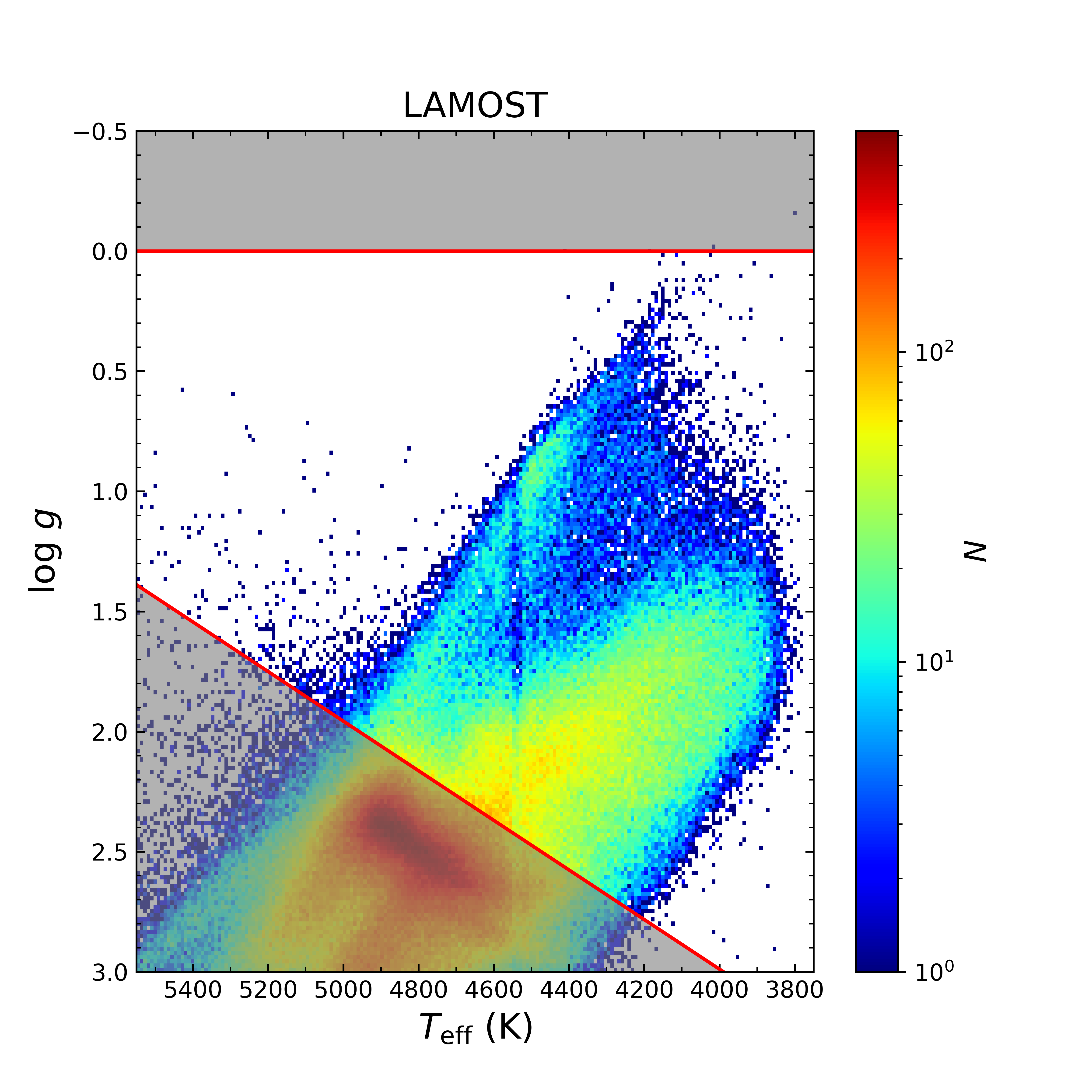}
\caption{The distribution of APO-LRGBs (left panel) and LM-LRGBs (right panel) on the $T_{\rm eff}$-log\,$g$ plane. The stars covered by the shadow regions are excluded from our LRGB samples.}
\end{center}
\label{dist}
\end{figure*}

\begin{figure}
\begin{center}
\includegraphics[width=0.45\textwidth]{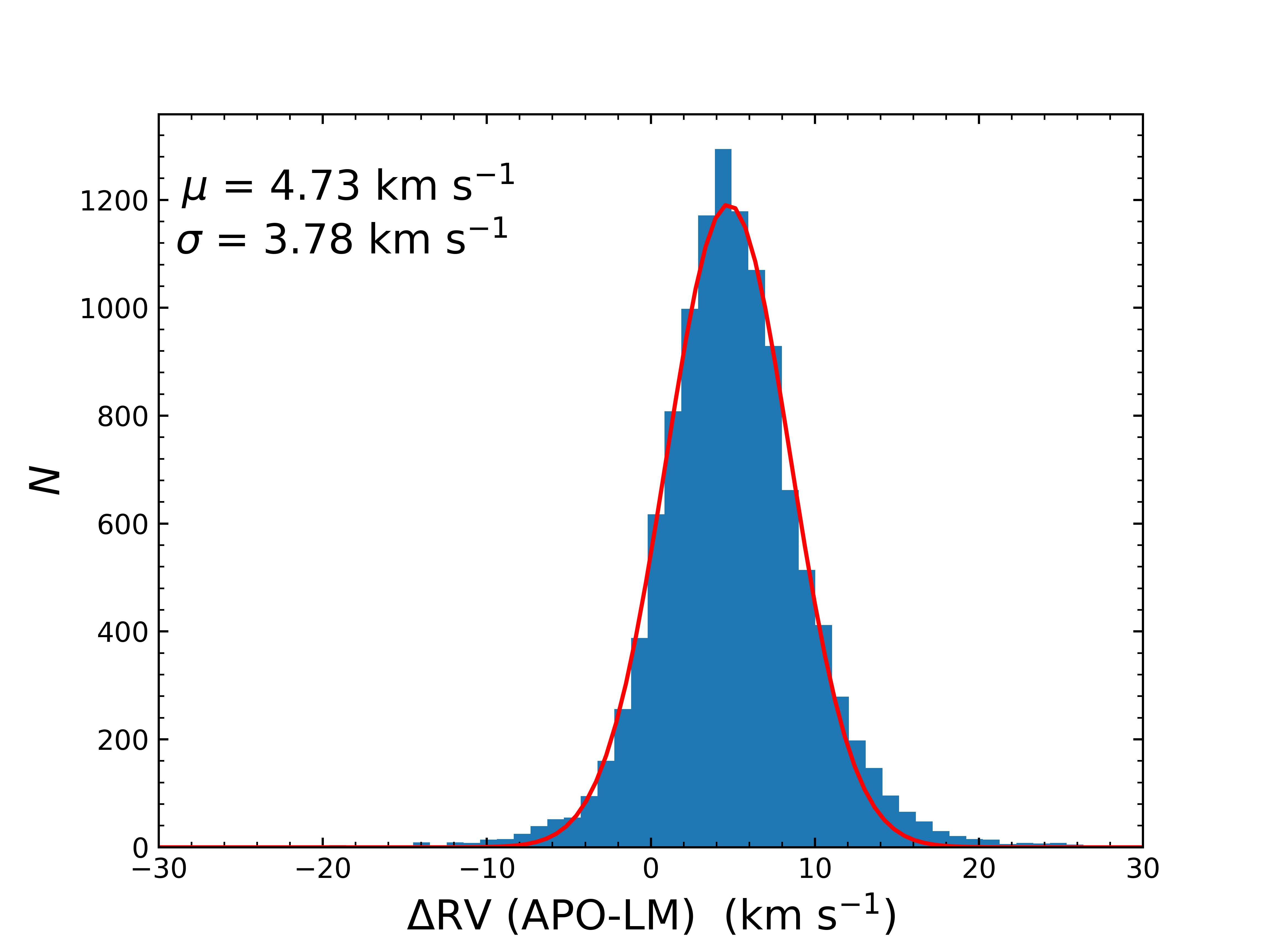}
\end{center}
\caption{Distribution of radial velocity difference between APO-LRGBs and LM-LRGBs. Overplotted in red is a Gaussian fit to the histogram distribution. The values of the mean and the standard deviation yielded by the fit are marked in the top-left corner. }
\label{RC}
\end{figure}

For each survey, those stars with highest SNRs are kept in the final catalog if they are observed greater than 2 times. In this way, 145,223 and 121,437 unique LRGBs are left in APOGEE (hereafter the APO-LRGB sample) and LAMOST (hereafter the LM-LRGB sample) surveys, respectively. The radial velocity (RV) is a key to determine the RC.
By using 11,778 common LRGBs, the zero-point of radial velocity measured by the two surveys are checked. As shown in Fig.\,\ref{RC}, a mean offset of $+4.73$\,km\,$^{-1}$ (APOGEE minus LAMOST) is detected with a small scatter of only 3.78\,km\,s$^{-1}$. The result is consistent with the independent check by \citet{2018AJ....156...90H}. As shown by \citet{2018AJ....156...90H}, the zero-point scale of APOGEE is close to that of the radial velocity standard stars. We thus apply a zero-point correction of $4.73$\,km\,$^{-1}$ to the radial velocity of LM-LRGBs. For common targets, the one from the APOGEE is kept in the final catalog, and finally 254,882 unique LRGB stars are found in the two surveys.  

\section{Distances of LRGB stars}
\subsection{Method}
\begin{figure}
\begin{center}
\includegraphics[width=0.45\textwidth]{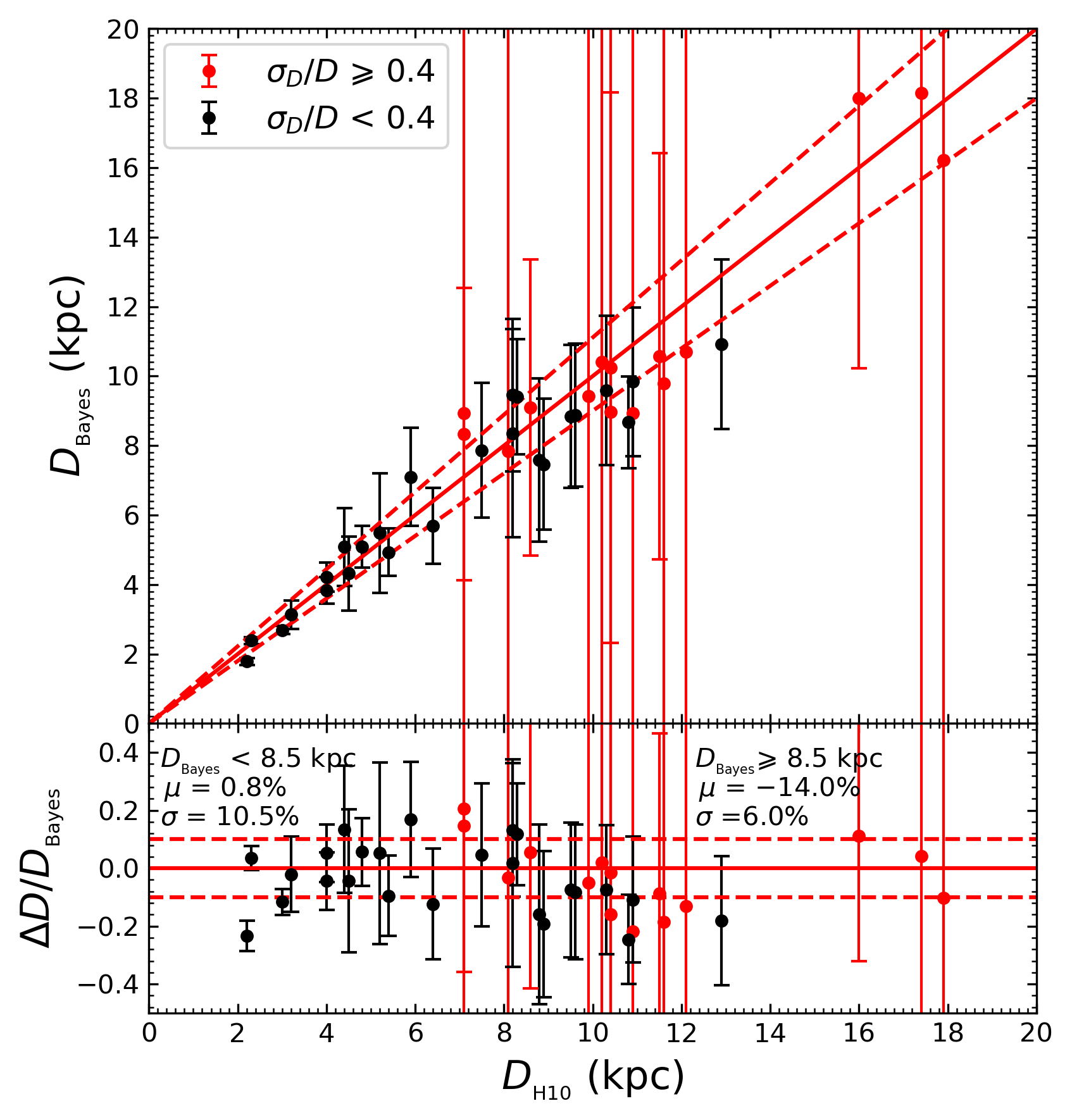}
\end{center}
\caption{Distances of GCs derived from the Gaia parallax measurements by a Bayesian approach compared to those from H10. The black and red dots indicate the mean distances of GCs with the relative distance error $\sigma_D/D <$ 0.4 and $\sigma_D/D \geqslant$ 0.4, respectively, estimated from the measurements of Gaia parallax. The red dashed lines mark $D_{\rm Bayes} = 0.9D_{\rm H10}$ and $D_{\rm Bayes} = 1.1D_{\rm H10}$. The ratios $\Delta D$/$D_{\rm_{Bayes}}$ are shown in the lower panel, with the mean and standard deviation of the relative difference $\Delta D/D_{\rm_{Bayes}}$ marked in the top corners. The dashed red lines in the lower panel indicate $\Delta D/D_{\rm_{Bayes}} = \pm0.10$.}
\label{li}
\end{figure}

In the present study, the distance estimates of the adopted LRGBs are the key to the construction of the Galactic RC. 
In principle, we can determine the distances of LRGBs by the parallax measurements of Gaia\,EDR3.
Doing so, we adopt a Bayesian approach \citep[see][for detail]{2021ApJ...907L..42H} to estimate the distance from parallax. 
Here, the parallax has been corrected for the zero points provided by \citet{2021A&A...649A...4L}. 
We validate the resulted distances by over two thousand member stars in 38 globular clusters (GCs; see Appendix B for details about the selection of GC member stars).
The result is presented in Fig.\,\ref{li} and one can clearly see the LRGB distances from parallax are consistent with those of GCs from \citet[][hereafter H10]{2010arXiv1012.3224H}, with a scatter of about 10 per cent and a negligible offset in general. 
However, for distant LRGBs (i.e., larger than 8.5\,kpc), the distances from parallaxes have suffered large random (due to the large parallax error measurements of Gaia EDR3, especially those GCs marked as red dots) and systematic errors. 
Even excluding those GCs with large random errors from parallax measurements, the yielded distances are still smaller than those of GCs (black dots in Fig.\,\ref{li}) by 14.0 per cent on average. 

To obtain reliable distance estimates of all our sample stars, a data-driven method (similar to that of H19) is developed to predict the absolute magnitudes of LRGBs from near-infrared/optical spectra. 
This method is based on the assumption that LRGBs are standardizable candles that their absolute magnitudes can be well predicted from the features extracted from the observations (i.e., spectroscopy and photometry).
In this paper, the gradient boosted decision trees \citep[GBDT;][]{12345, Friedman2001GreedyFA} data-driven method is adopted. The GBDT regressors, using decision trees as weak learners, can be considered as an additive model in the following form: 
\begin{equation}
F(x) = \sum_{m=1}^{M} \nu h_m(x),
\end{equation}
where $h_m(x)$ is the basic function and it is often called as weak learners in the gradient boosting algorithm scenario, $x$ is the given input, $m$ is the number of trees and the parameter $\nu$ is called the learning rate. As a boosting technique, the GBDT is operated in a greedy manner,

\begin{equation}
F_m(x)=F_{m-1}(x)+\nu h_m(x).
\end{equation}

In each stage of training, the newly added tree $h_m$, given the previous ensemble $F_{m-1}$, is fitted in order to minimize a sum of losses $L_m$, i.e.,

\begin{equation}
h_m= \arg\min_{h}L_m =\arg\min_{h} \sum_{i=1}^{n} l(y_i, F_{m-1}(x_i) + \nu h(x_i)),
\end{equation}
where $ l(y_i, F(x_i))$ is defined by the so-called loss parameter and $h_m$ is then fitted to the predict value given by the negative gradients of the samples.

For our GBDT algorithm, we need to specify the parameters of the model at the beginning of training. Here, the number of trees ($n\_estimators$) we adopted is 500, the maximum depth of the tree ($n\_depth$) is 10 and other parameters are set to $learning\_rate$ = 0.1, $min\_samples\_split$ = 6, $min\_samples\_leaf$ = 4, $max\_leaf\_nodes$ = 4. The training process is carried out by using the scikit-learn package \citep{Pedregosa2011ScikitlearnML} in Python.

Before the training, the input spectra should be pre-processed as continuum-normalized spectra. We use the APOGEE spectra of $\lambda$\,15160-15780\,\AA \,, $\lambda$\,15890-16300\,\AA \,\,and $\lambda$\,16490-16900\,\AA \, and the LAMOST spectra of $\lambda$\,3900-5800\,\AA \,\,and $\lambda$\,8450-8950\,\AA . For APOGEE spectra, the  wavelength windows are chosen to avoid potential artificial effects \citep[e.g., wing-like features mentioned in][]{2021AJ....161..167A} in the edges of the red, green and blue spectrographs. For the LAMOST spectra, the wavelength windows are chosen to include the most informative parts \citep[e.g.][]{2022AJ....163..149W}. All spectra are normalized by dividing their pseudocontinuum spectra. Doing so, each spectrum is smoothed by the median filtering method with a filter size of 21 pixels. Then we use fourth and fifth-order polynomial for APOGEE and LAMOST to fit the smoothed spectra to define the pseudocontinuum spectra. For each star $n$ in our samples, we create a $D$-dimensional feature vector $x_n$,
\begin{center}
$x_n^T =[f_{1n},f_{2n},f_{3n},... ,f_{kn}]$,
\end{center}where the $x_n^T$ includes 6,522-dimensional features from spectra for APO-LRGBs, and 2,400-dimensional features from spectra for LM-LRGBs respectively. 

The key to train precise relations between absolute magnitudes and spectral features is to carefully define training and testing samples with accurate absolute magnitudes determined and wide coverages in stellar parameters (e.g., effective temperature, metallicity and luminosity) and in spatial distributions. 
The later requirement is very important for unbiased absolute magnitude predictions given the large spatial and parameter coverage of the LRGB stars. 
Doing so, our training and testing samples consist of stars selected from different criteria: 1) stars with precise distances estimated from Gaia parallax measurements in low-extinction regions (14\% systematics has been corrected for those distant ones with $d \ge 8.5$\,kpc as found in Fig.\,\ref{li}); 2) distant halo stars and 3) distant disk stars. The details are described in Appendix\,B. In total, 10,145 stars, with 7,101 stars as the training set and 3,044 stars as the testing set, are selected for APO-LRGB stars. 
For LM-LRGB stars, 9,063 stars (6,484 stars as the training sample and 2,779 stars as the testing sample) are selected.

\begin{figure*}
\begin{center}
\includegraphics[width=0.45\textwidth]{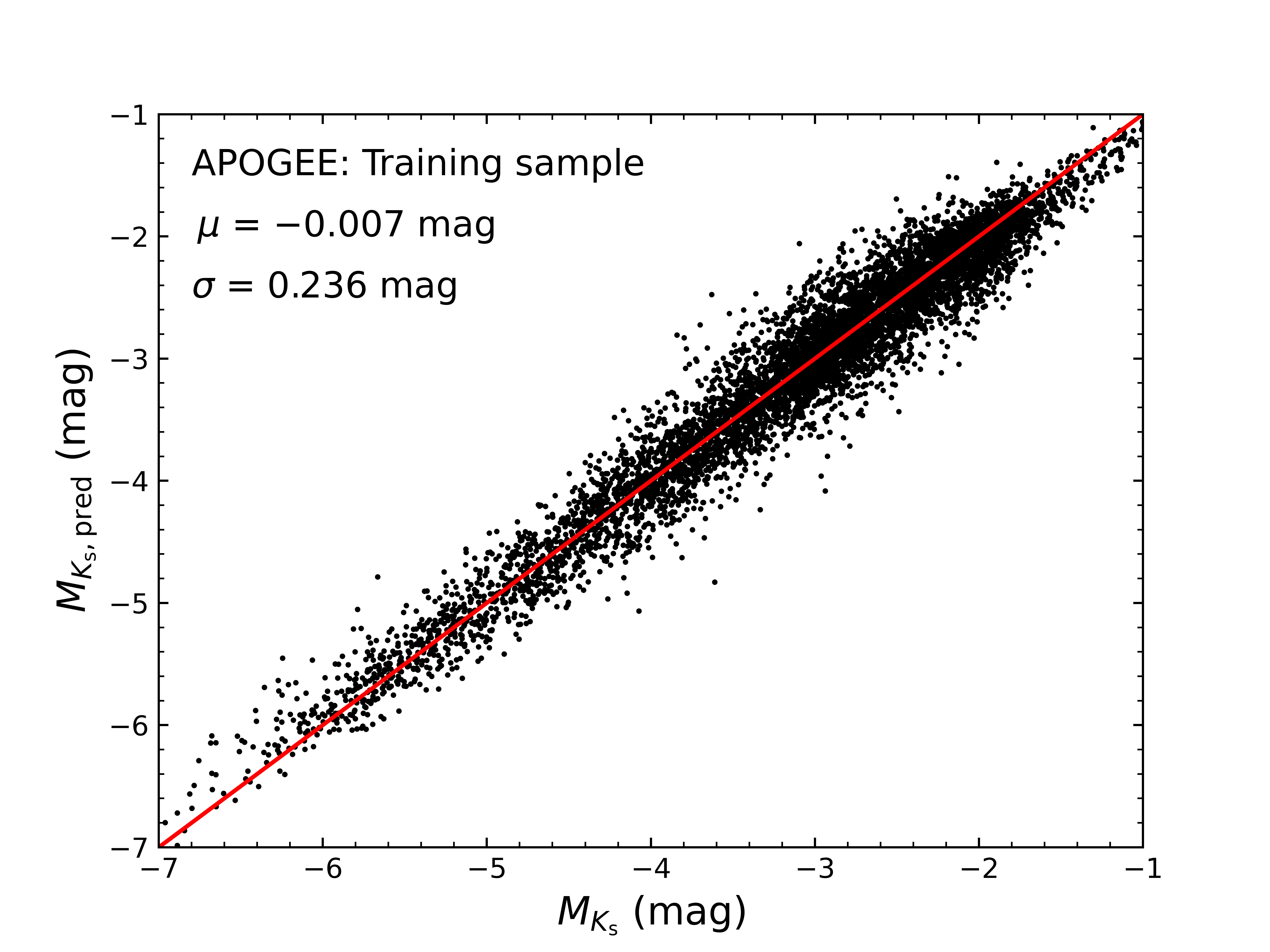}
\includegraphics[width=0.45\textwidth]{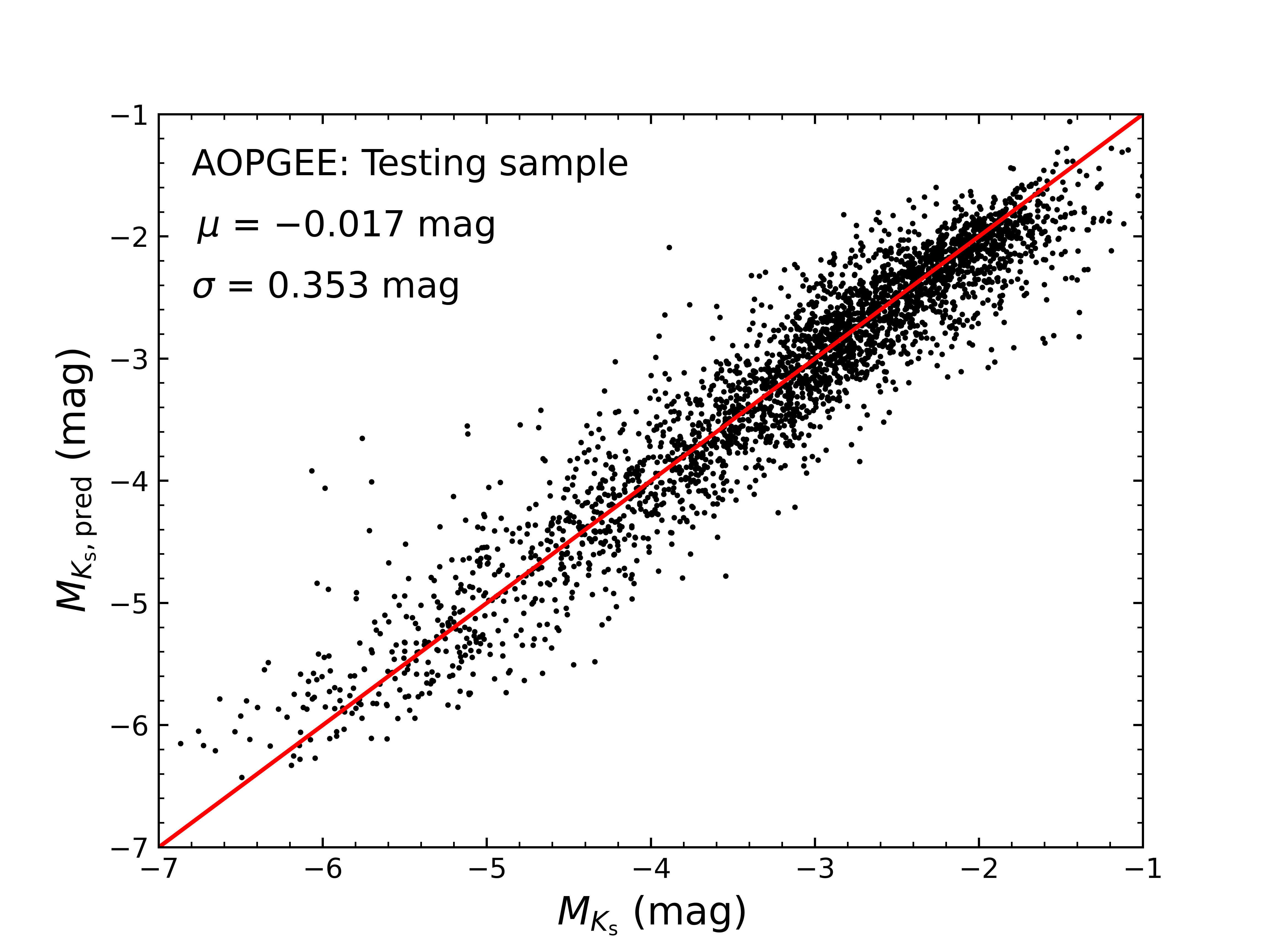}
\includegraphics[width=0.45\textwidth]{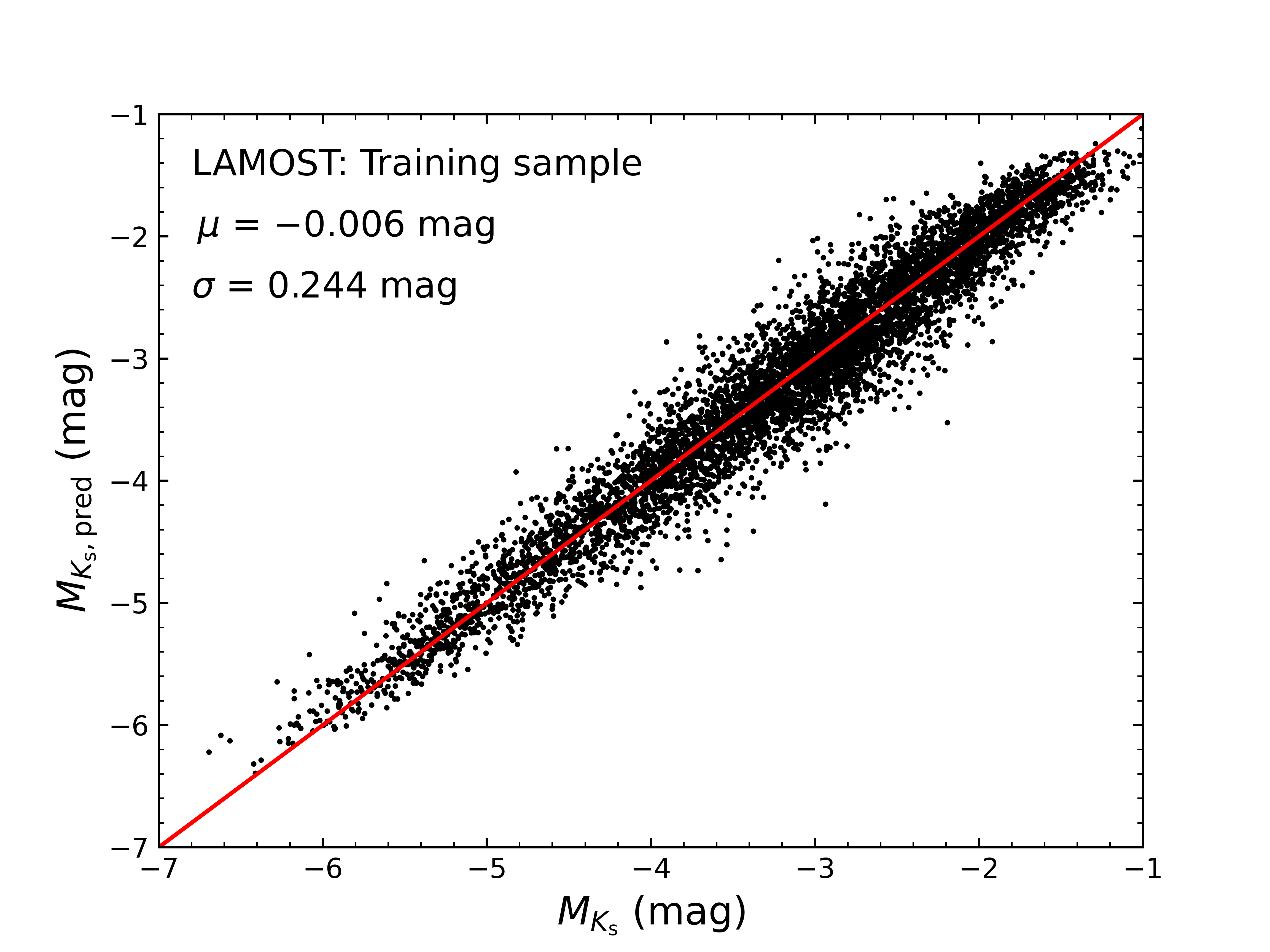}
\includegraphics[width=0.45\textwidth]{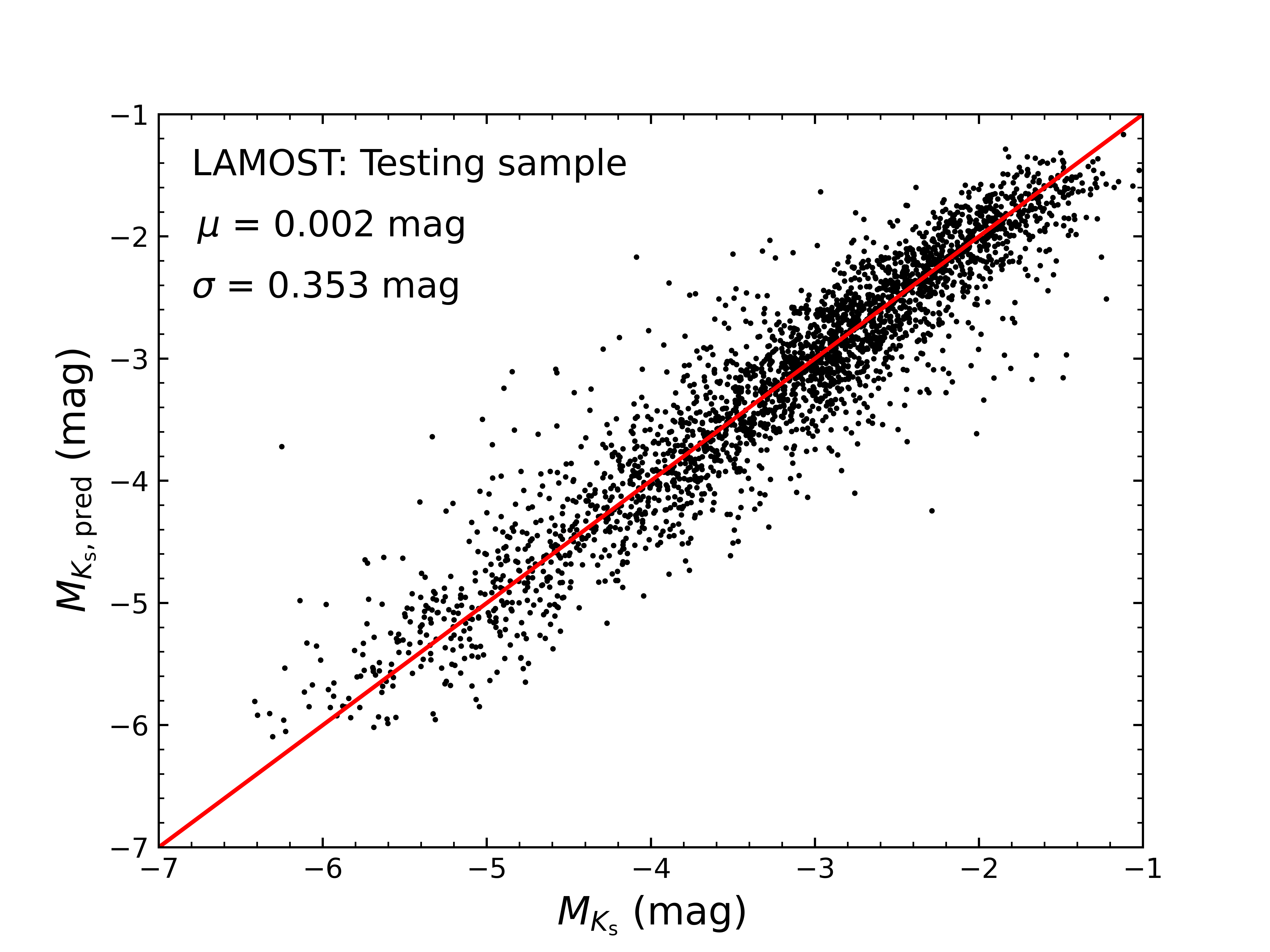}
\end{center}
\caption{Comparisons of input absolute magnitudes $M_{K_{\rm s}}$ and predicted ones $M_{K_{\rm s},{\rm pred}}$ from APOGEE (top panel)/LAMOST (bottom panel) spectra for the training (left panel) and the testing samples (right panel). The systematic differences, as well as the standard deviations, are marked in the top corners.}
\label{56}
\end{figure*}

With the training set defined above, the relations between $D$-dimensional feature vector $x_n$ and $K_{\rm s}$ band absolute magnitudes are constructed based on the GBDT. Here, the absolute magnitudes of $K_{\rm s}$ band are chosen due to the low extinction suffered in this band. The training and testing results for the absolute magnitudes of $K_{\rm s}$ are shown in Fig.\,\ref{56}. The systematic differences between input absolute magnitudes and absolute magnitudes derived from APOGEE/LAMOST spectra are tiny with $\Delta M_{K_{\rm s}} \sim$\,$-$0.007/$-$0.006\,mag for the training samples and $\Delta M_{K_{\rm s}} \sim$\,$-$0.017/0.002\,mag for the testing samples. The precisions are typically 0.35\,mag found by the testing samples, slightly larger than the value of 0.24\,mag achieved by the training results, implying a precision of 10-16\% achievable for the derived distances. We then apply the constructed relations to all the APO-LRGB and LM-LRGB sample stars to derive the $K_{\rm s}$ band absolute magnitudes. The distances are for those stars are further derived by combinations of the apparent and absolute $K_{\rm s}$ band magnitudes, as well as the extinction values from the ``star pair" technique (see Appendix\,A).

\begin{figure*}
\begin{center}
\includegraphics[width=0.4\textwidth]{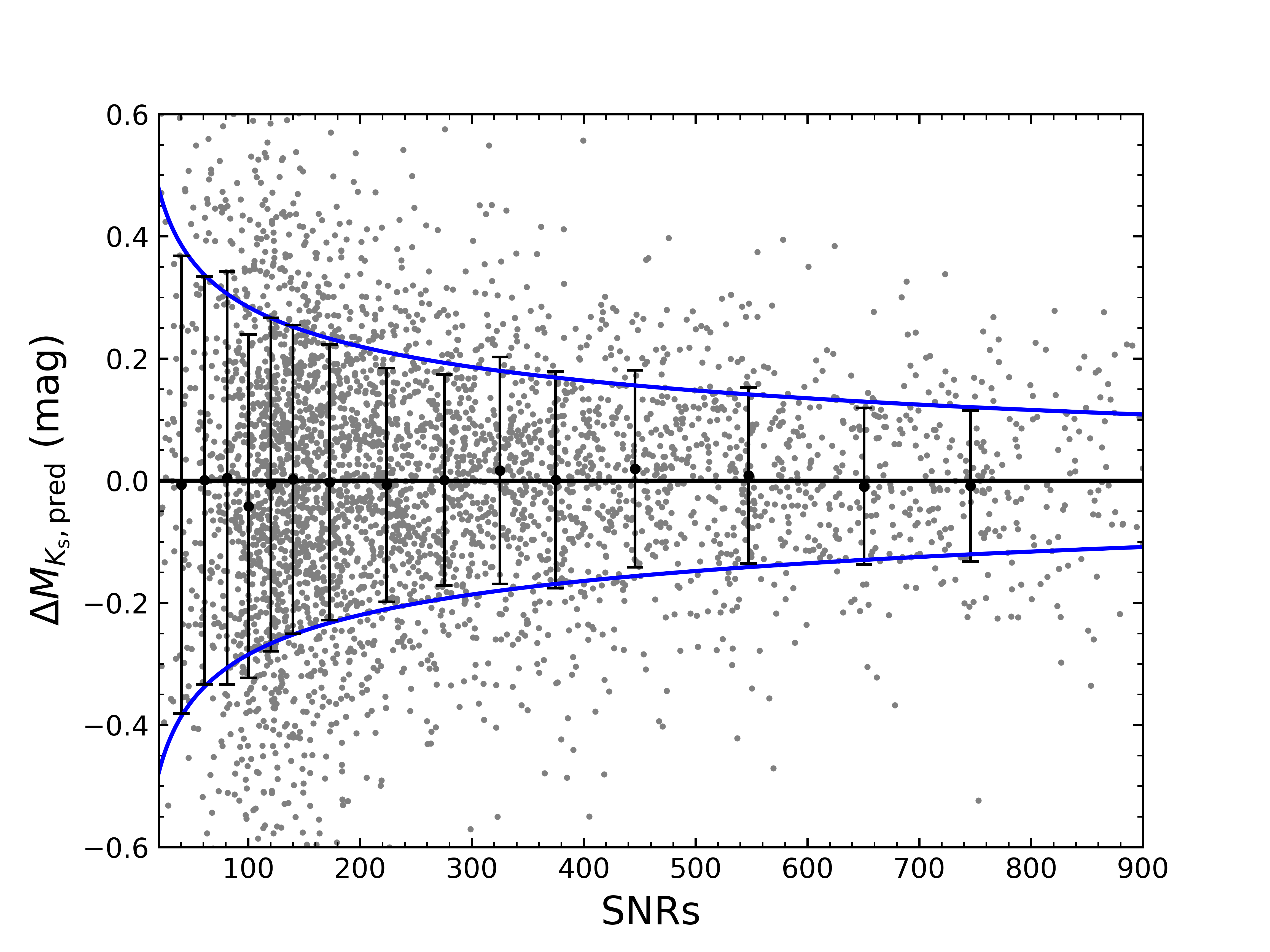}
\includegraphics[width=0.4\textwidth]{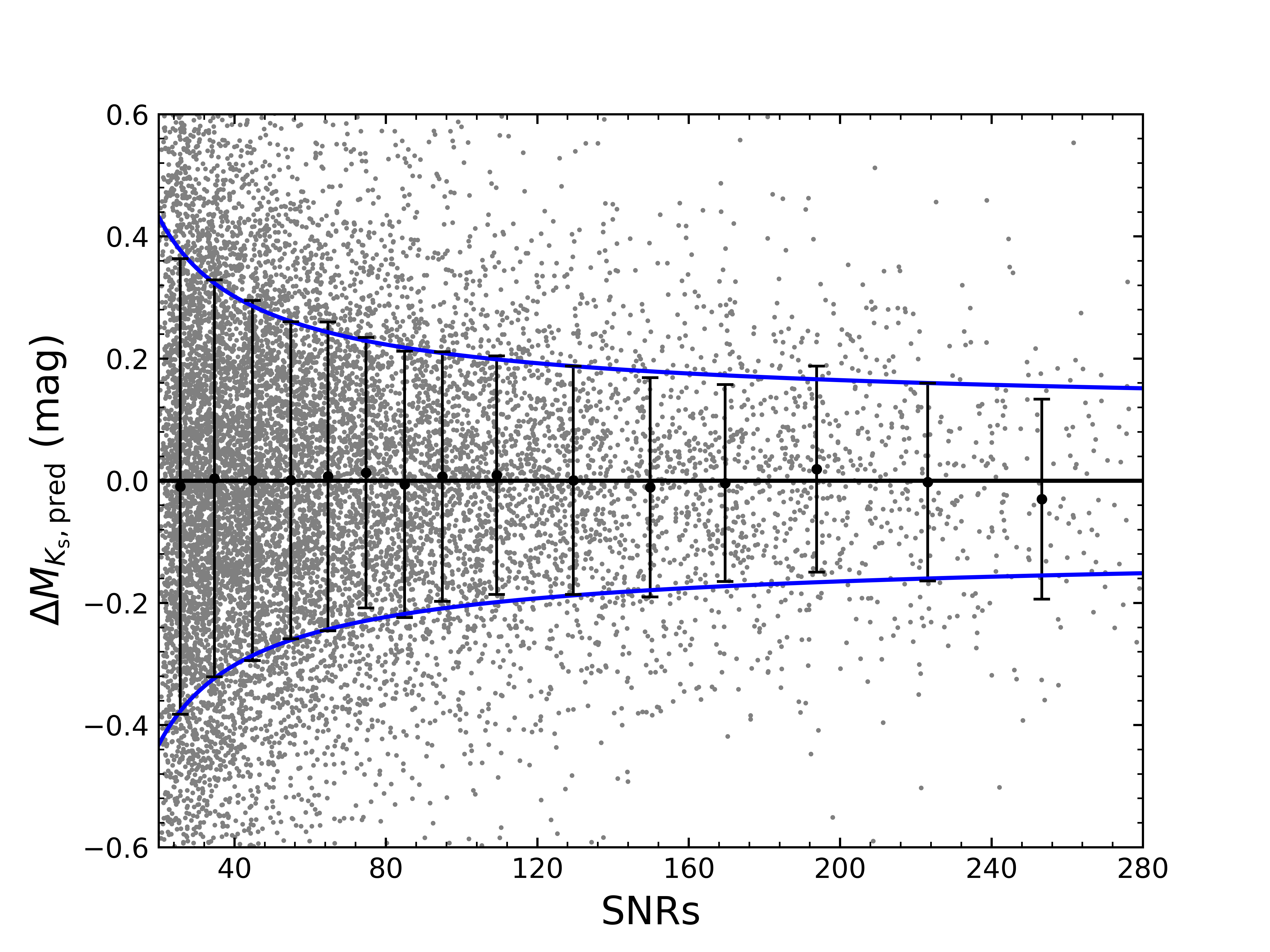}
\end{center}
\caption{Left panel: Internal residuals of $K_{\rm s}$ band absolute magnitudes for APO-LRGBs given by duplicate observations of similar spectral SNRs at different spectral SNRs bins. Gray dots are the differences of magnitudes predicted from duplicate observations of SNRs differences smaller than 20\%. Black dots and error bars represent the medians and standard deviations (after divided by $\sqrt2$) of the differences in the individual spectral SNRs bins. Blue lines indicate fits of the standard deviations as a function of spectral SNRs. Right panel: same as left plane but for LM-LRGBs.}
\label{snrdif}
\end{figure*}

\subsection{Uncertainties of the predicted magnitudes}

In this section, we examine the uncertainties of the predicted magnitudes from the relations trained by the GBDT. Generally, the uncertainties of these predicted magnitudes depend on spectral noise (random errors) and method errors. In order to evaluate the random errors of the predicted magnitudes of the LRGB stars, repeated observations of similar spectral SNRs (with differences less than 20\%) collected at different nights were used. Fig.\,\ref{snrdif} shows the predicted magnitude differences between duplicated observations (divided by $\sqrt{2}$) as a function of the mean spectral SNRs. The standard deviations of the magnitude differences show significant decrease trend with increasing spectral SNRs. To assign a proper random error for each target, we fit the dispersions of the magnitude differences by adopting the functions used by \citet{2020ApJS..249...29H}:
\begin{equation}
\sigma_{\rm{r}}= a+\frac{c}{({\rm{SNR}})^b},
\end{equation}
where $\sigma_{\rm{r}}$ represents the random error. The resulting fit coefficients for the APO-LRGB and LM-LRGB samples are presented in Table\,\ref{aaaa}. The method errors $\sigma_{m}$ used here are the typical standard deviations of the training results found in above section, i.e., 0.236 for APO-LRGB sample and 0.244 for LM-LRGB sample, respectively. The final error of each program star is then given by $\sqrt{\sigma_{\rm{r}}^2 + \sigma_{\rm{m}}^2}$. The uncertainties are mainly contributed by the random errors and method errors, respectively, at low and high spectral SNRs.

\begin{table}
\centering
\caption{The Fit Coefficients of Random Errors of $K_{\rm s}$ band absolute Magnitude}
\setlength{\tabcolsep}{3mm}{
\begin{tabular}{ccccc}
\hline
\hline
Sample & $a$ &$ b $& $c$ \\
\hline
APO-LRGBs & $-$0.006& 0.406 &1.935\\ 
LM-LRGBs & 0.008& 0.643 & 2.341\\
\hline
\end{tabular}}
\label{aaaa}
\end{table}

\begin{figure*}
\begin{center}
\includegraphics[width=0.45\textwidth]{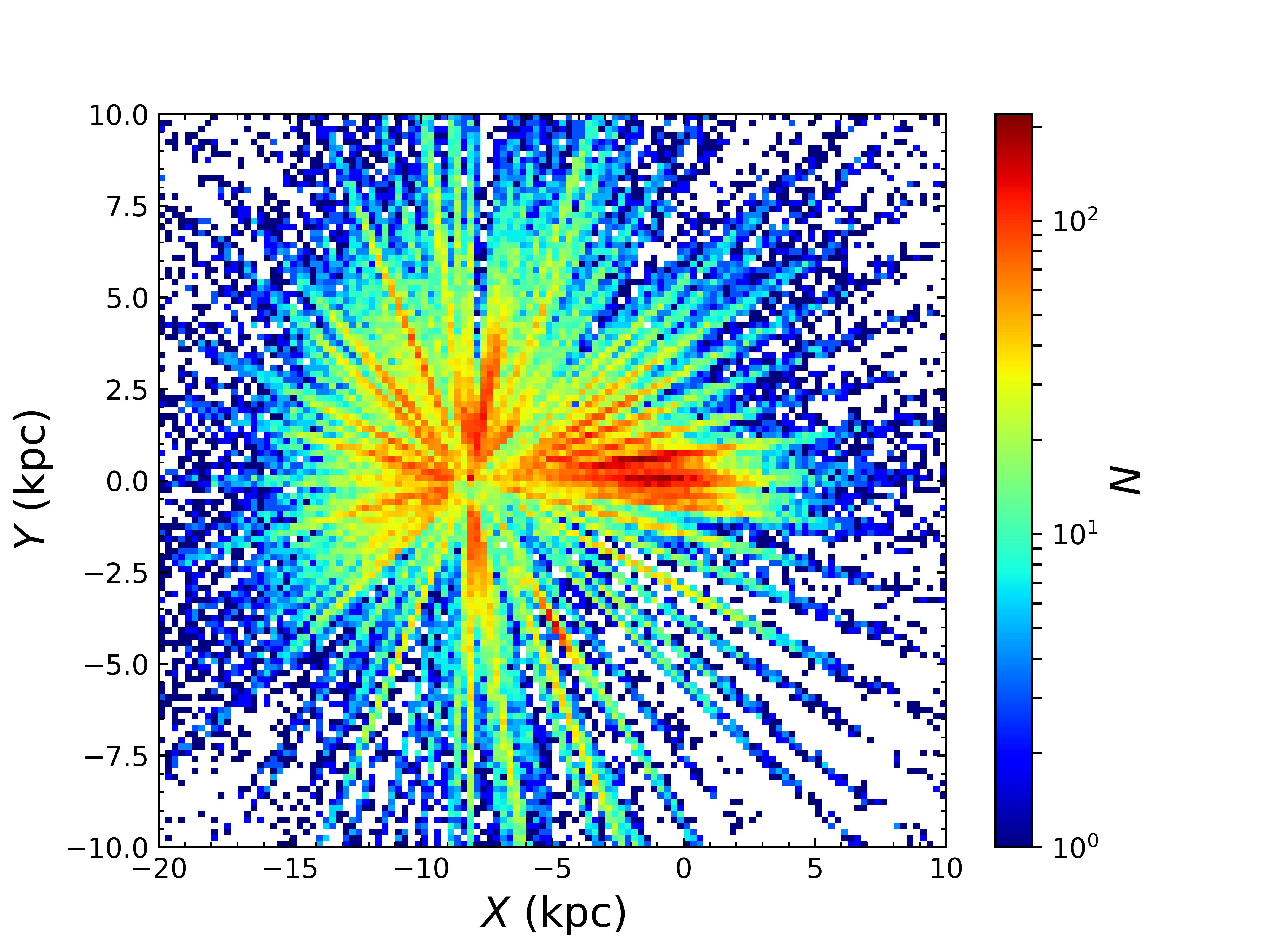}
\includegraphics[width=0.45\textwidth]{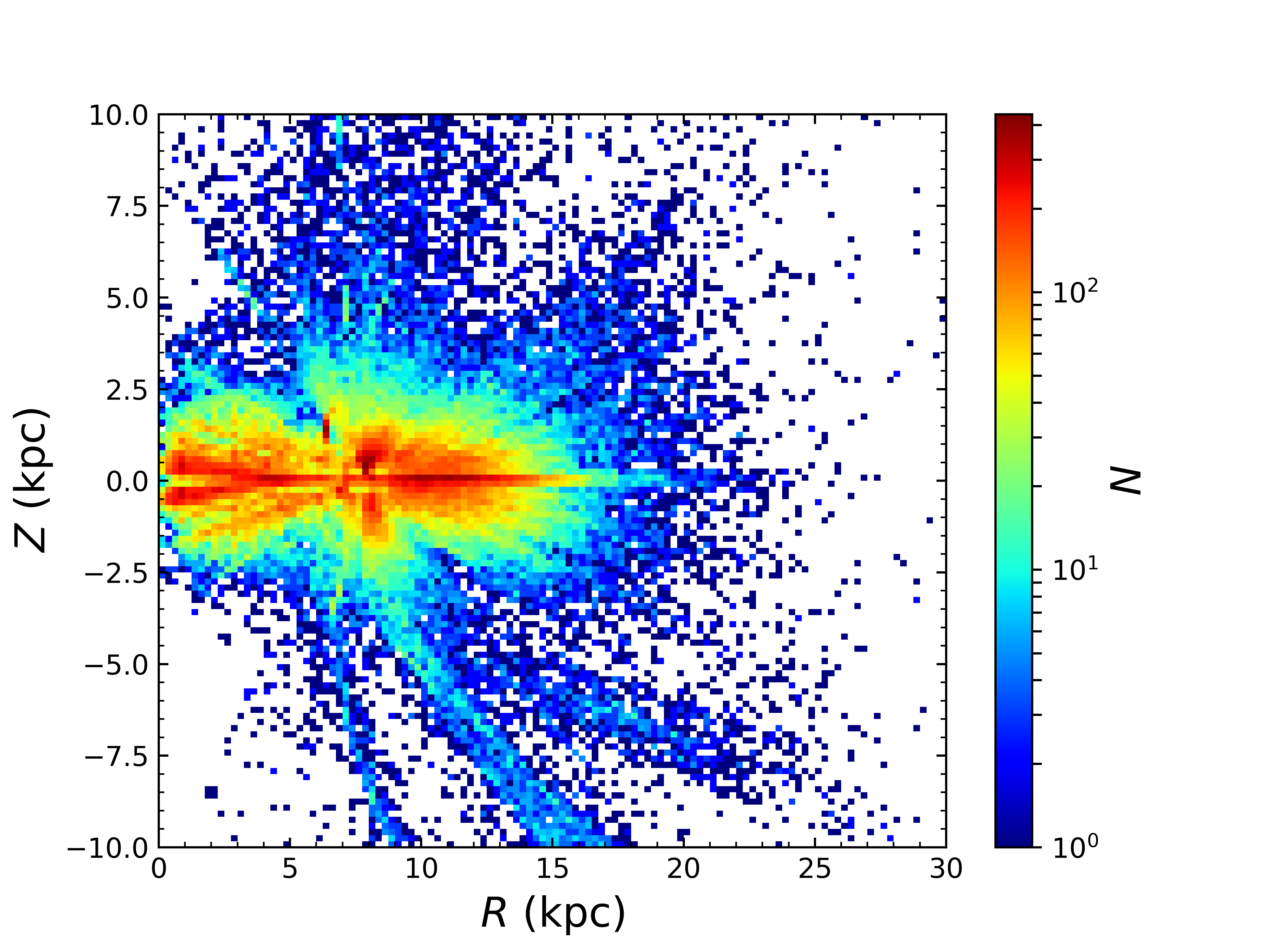}
\end{center}

\begin{center}
\includegraphics[width=0.45\textwidth]{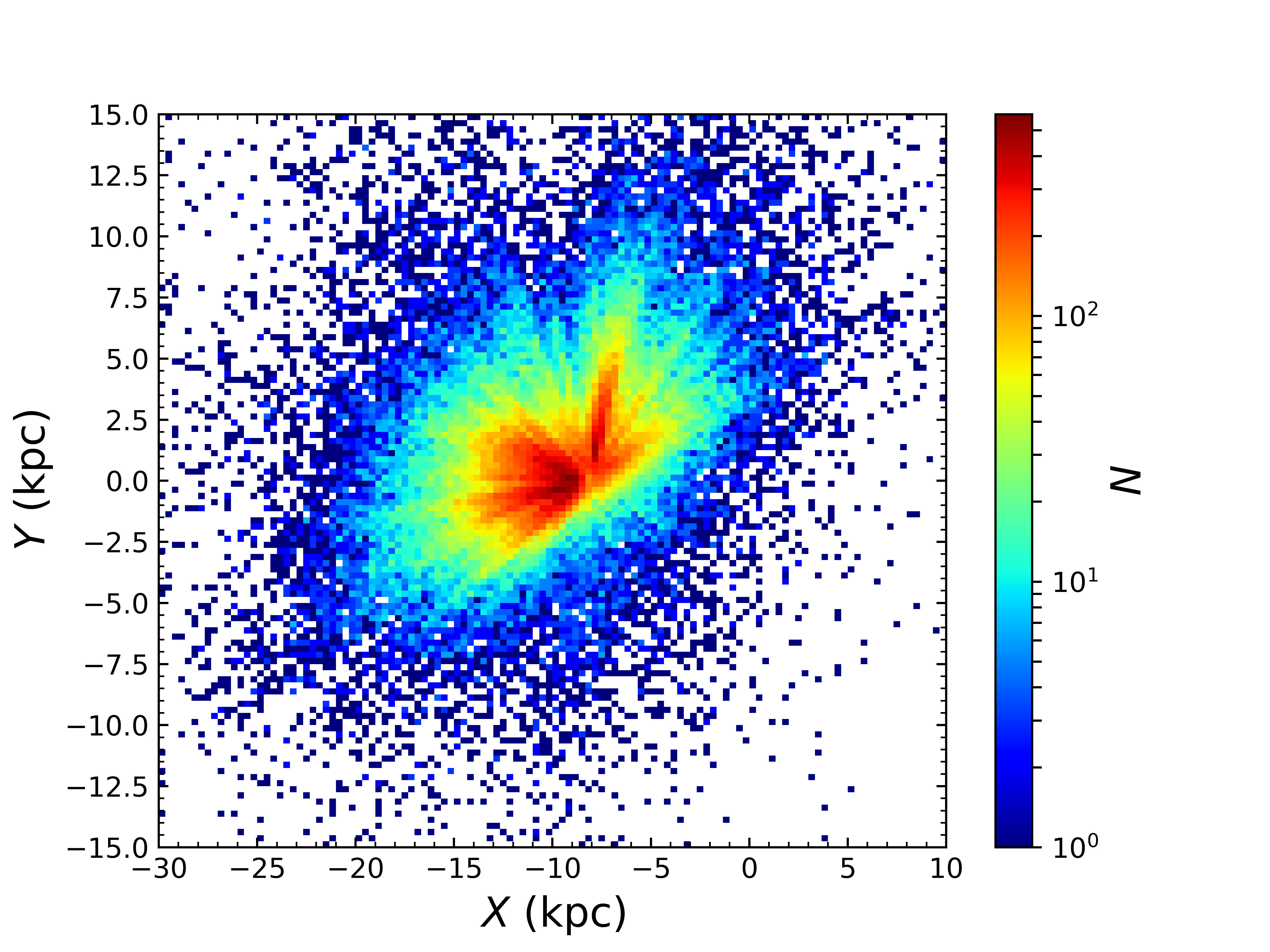}
\includegraphics[width=0.45\textwidth]{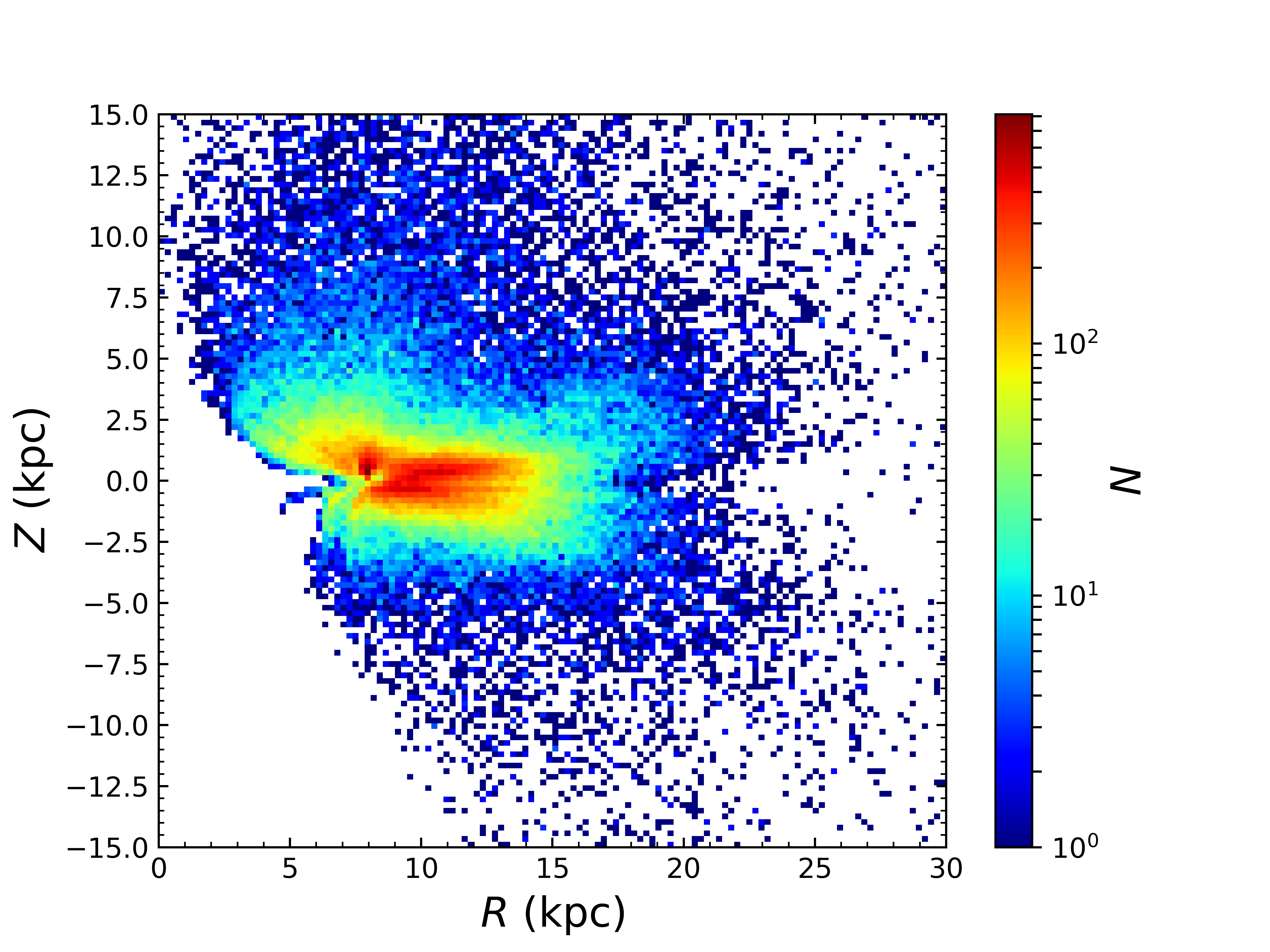}
\end{center}
\caption{Number density of APOGEE (top panels) and LAMOST (bottom panels) LRGB sample stars on the $X$ $-$ $Y$ (left panels) and $R$ $-$ $Z$ (right panels) planes.}
\label{zzz}
\end{figure*}

We then calculated the 3D positions for our sample stars by their sky positions $(l, b)$ and measured distances. The $X$ $-$ $Z$ and $R$ $-$ $Z$ plane distributions of the APO-LRGB and LM-LRGB sample stars are presented in Fig.\,\ref{zzz}. 
Our final LRGB sample covers a large volume both of the Galactic disk(s) and halo, with $R$ (the projected Galactocentric distance in a Galactocentric cylindrical system) ranging from the Galactic center to 30\,kpc, and $|Z| \leqslant 15$\,kpc.

\subsection{Validation of distance estimates}

\begin{figure}
\begin{center}
\includegraphics[width=0.4\textwidth]{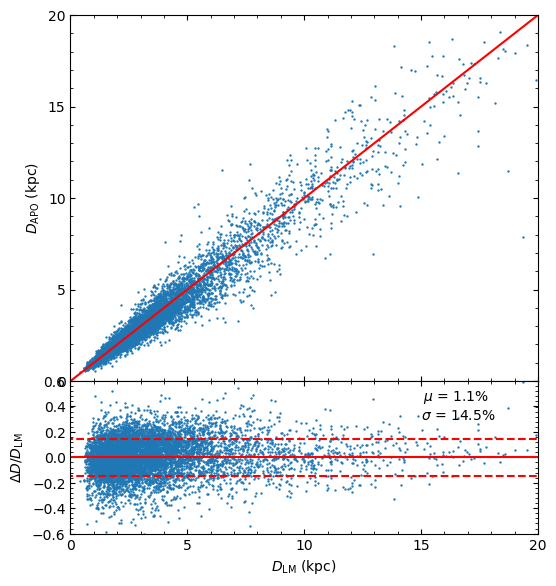}
\end{center}
\caption{Comparison of estimated distances between the common stars of APO-LRGBs and LM-LRGBs. The ratios $\Delta D$/$D_{\rm{LM}}$ are shown in the lower panel, with the median and standard deviation of the relative distance difference $\Delta D$/$D_{\rm{LM}}$ marked in the top-right corner. The red dashed lines represent the 1-$\sigma$ scatters.}
\label{V}
\end{figure}

As mentioned in Section\,2.3, in addition to evaluate the random errors, common sources also can be used to verify whether there is a systematic difference in the distance between the APO-LRGB and LM-LRGB samples. Fig.\,\ref{V} clearly shows that, the median deviation of the relative distance differences of the common sources is negligible with only of 1.1\% (APOGEE minus LAMOST). The dispersion of the relative distance differences is 14.5 per cent, in consistent with the result reported by the training samples.

\begin{figure}
\begin{center}
\includegraphics[width=0.4\textwidth]{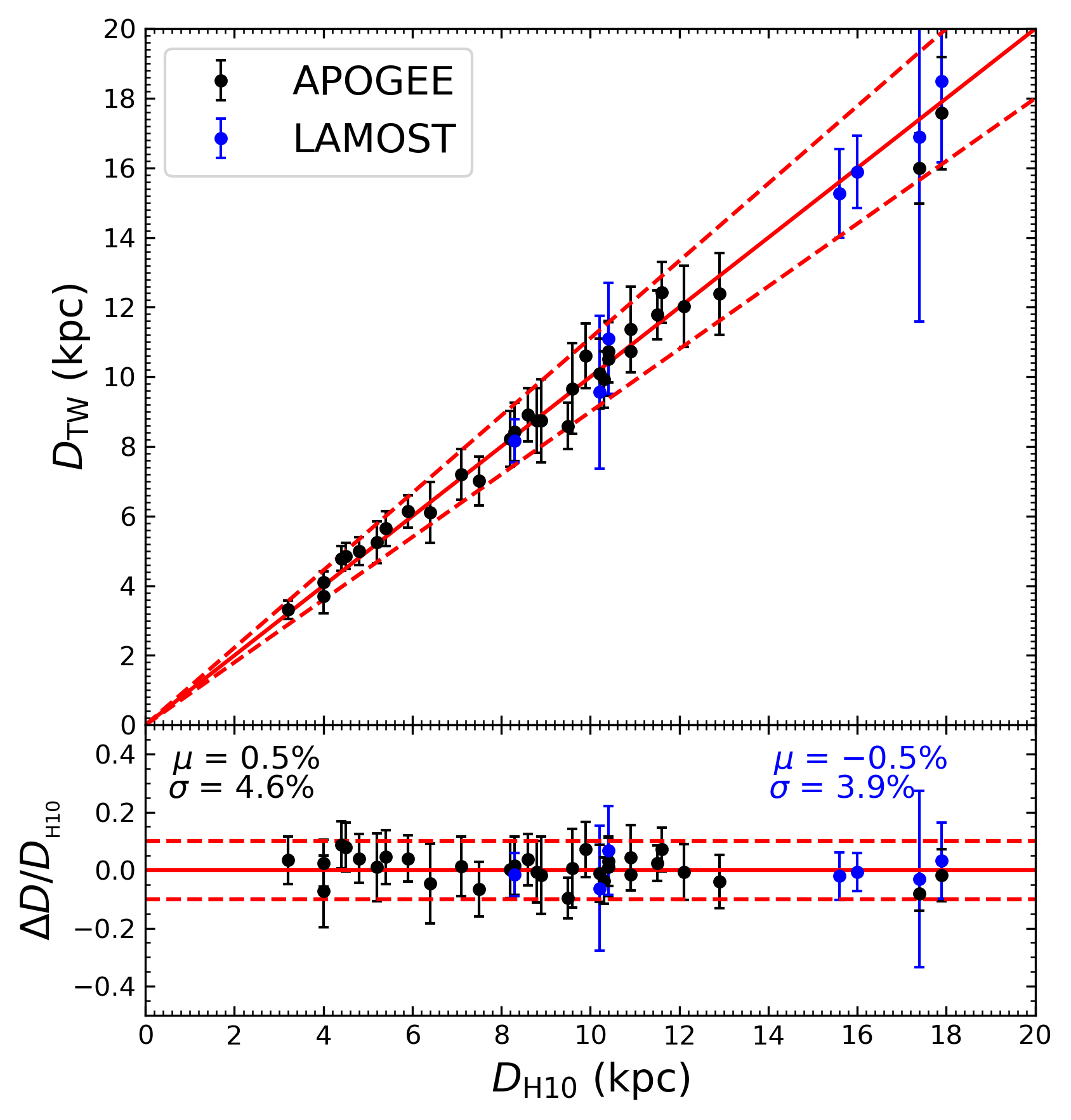}
\end{center}
\caption{Comparison of estimated distances of GCs with those from H10. The black and blue dots indicate the mean distances of GC members estimated from APO-LRGBs and LM-LRGBs, respectively. The red dashed lines mark $D_{\rm TW} = 0.9D_{\rm{H10}}$ and $D_{\rm TW} = 1.1D_{\rm{H10}}$. The ratios of $\Delta D$/$D_{\rm{H10}}$ are shown in the lower panel, with the mean and standard deviation of the relative distance difference $\Delta D/D_{\rm{H10}}$ marked in the top corners. The dashed red lines in the lower panel indicate $\Delta D/D_{\rm_{H10}} = \pm0.10$.}
\label{gc}
\end{figure}

To provide external checks of our spectrophotometric distances predicted from spectra by using the relations constructed by the GBDT, we select 780/55 stars (not used by the training samples) from APO-LRGB/LM-LRGB sample belonging to 32/7 GCs to validate the accuracy of the estimated distances. As shown in Fig.\,\ref{gc} and Table\,2, we compare the spectrophotometric distance (hereafter $D_{\rm TW}$) to distances of GCs taken from H10 (hereafter $D_{\rm H10}$). 
Our spectrophotometric distances of GCs are in excellent agreement with the distances of H10, with a negligible median offset of the relative distance difference $\Delta D/D$ of 0.5\%/$-$0.5\% and a standard deviation of 4.6\%/3.9\% for APO-LRGBs/LM-LRGBs, respectively. 
Meanwhile, the mean dispersions of the member distances revealed by individual GCs are 10-15 per cent on average, compared to the distances of the clusters, both for APO-LRGB and LM-LRGB samples, in agree with the results found for the training and testing samples.

\begin{table*}
\centering
\begin{threeparttable}
\caption{Comparison of our distances of GCs with the values of $D_{\rm H10}$.}
\setlength{\tabcolsep}{2.5mm}{
\begin{tabular}{cccccccc}
\hline
\hline
Name & RA & DEC & RV & $D_{\rm{H10}}$ & $D_{\rm{TW}}$ &$N$ \\
& (degree) & (degree) & (km\,s$^{-1}$) & (kpc) & (kpc)& & \\
\hline
&  &  & (a)\,APO-LRGBs  & & &  \\
\hline
 BH 229&262.772&$-$29.982&40.61& 8.2&         8.22$\pm$0.81&10\\
NGC 104&6.024&$-$72.081&$-$17.21&      4.5&   4.86$\pm$0.38&16\\
 NGC 288&13.188&$-$26.583&$-$44.83&   8.9&    8.74$\pm$1.19&22\\
 NGC 362&15.809&$-$70.849&223.26&     8.6&    8.91$\pm$0.76&27\\
 NGC 1851&78.528&$-$40.047&320.30&     12.1&    12.03$\pm$1.16&22\\
 NGC 1904&81.046&$-$24.525&205.84&    12.9&    12.39$\pm$1.18&25\\
 NGC 2808&138.013&$-$64.864&103.90&    9.6&    9.67$\pm$1.30&39\\
 NGC 4590&189.867&$-$26.744&$-$92.99&   10.3&  9.93$\pm$0.81&21\\
 NGC 5024&198.230&18.168&$-$62.85&     17.9&    17.57$\pm$1.61&22\\
 NGC 5053&199.113&17.700&42.77&         17.4&    15.99$\pm$1.01&11\\
 NGC 5139&201.697&$-$47.480&234.28&     5.2&   5.25$\pm$0.61&85\\
 NGC 5272&205.548&28.377&$-$147.28&  10.2&     10.09$\pm$1.00&64\\
 NGC 5904&229.638&2.081&53.70&        7.5&     7.02$\pm$0.70&45\\
 NGC 6171&248.133&$-$13.054&$-$34.68& 6.4&    6.11$\pm$0.88&10\\
 NGC 6205&250.422&36.460&$-$244.49&    7.1&    7.20$\pm$0.73&22\\
 NGC 6218&251.809&$-$1.949&$-$41.35&   4.8&   4.99$\pm$0.40&16\\
 NGC 6254&254.288&$-$4.100&74.02&        4.4&   4.79$\pm$0.36&20\\
 NGC 6273&255.658&$-$26.268&145.54&    8.8&   8.75$\pm$0.93&24\\
 NGC 6293&257.543&$-$26.582&$-$143.66&  9.5&  8.59$\pm$0.66&11\\
 NGC 6304&258.634&$-$29.462&$-$108.62&   5.9& 6.14$\pm$0.47&15\\
 NGC 6316&259.155&$-$28.140&99.81&     10.4&    10.73$\pm$0.89&15\\
 NGC 6341&259.281&43.136&$-$120.48&   8.3&    8.43$\pm$0.84&40\\
 NGC 6380&263.617&$-$39.069&$-$6.54&   10.9&   10.74$\pm$0.60&14\\
 NGC 6388&264.072&$-$44.736&82.85&   9.9&     10.61$\pm$0.94&16\\
 NGC 6441&267.554&$-$37.051&17.27&   11.6&     12.43$\pm$0.88&10\\
 NGC 6569&273.412&$-$31.827&$-$49.83&  10.9&   11.37$\pm$1.23&10\\
 NGC 6656&279.100&$-$23.905&$-$147.76&  3.2&    3.31$\pm$0.26&23\\
 NGC 6752&287.717&$-$59.985&$-$26.28&  4.0&   3.71$\pm$0.50&17\\
 NGC 6809&294.999&$-$30.965&174.40&     5.4&   5.64$\pm$0.50&25\\
 NGC 6838&298.444&18.779&$-$22.27&   4.0&     4.10$\pm$0.32&18\\
 NGC 7078&322.493&12.167&$-$106.76&   10.4&    10.51$\pm$1.05&54\\
 NGC 7089&323.363&$-$0.810&$-$3.72&     11.5&   11.78$\pm$0.71&11\\

\hline
&  &   &(b)\,LM-LRGBs & & &  \\
\hline
NGC 5024&198.230&18.168&$-$62.85&    17.9&      18.50$\pm$2.34&18\\
 NGC 5053&199.113&17.700&42.77&      17.4&       16.89$\pm$5.29&6\\
 NGC 5272&205.548&28.377&$-$147.28&   10.2&    9.56$\pm$2.19&11\\
 NGC 5466&211.364&28.534&106.93&     16.0&     15.89$\pm$1.05&5\\
 NGC 6341&259.281&43.136&$-$120.48&   8.3&    8.17$\pm$0.62&5\\
 NGC 6934&308.547&7.404&$-$406.22&   15.6&     15.28$\pm$1.28&3\\
 NGC 7078&322.493&12.167&$-$106.76&     10.4&  11.11$\pm$1.59&7\\
 
\hline
\end{tabular}}
\begin{tablenotes} 
\item Col.\,1 is name of the GC, Cols.\,2$-$5 give the central right ascension, declination, mean radial velocity and distances of the GCs from H10, Cols.\,6 presents the mean value of our distance and it error, and Col.\,7 gives the number of GC member stars from the (a)\,APO-LRGBs or (b)\,LM-LRGBs.
\end{tablenotes} 
\end{threeparttable} 
\label{gcc}
\end{table*}

\begin{figure}
\begin{center}
\includegraphics[width=0.4\textwidth]{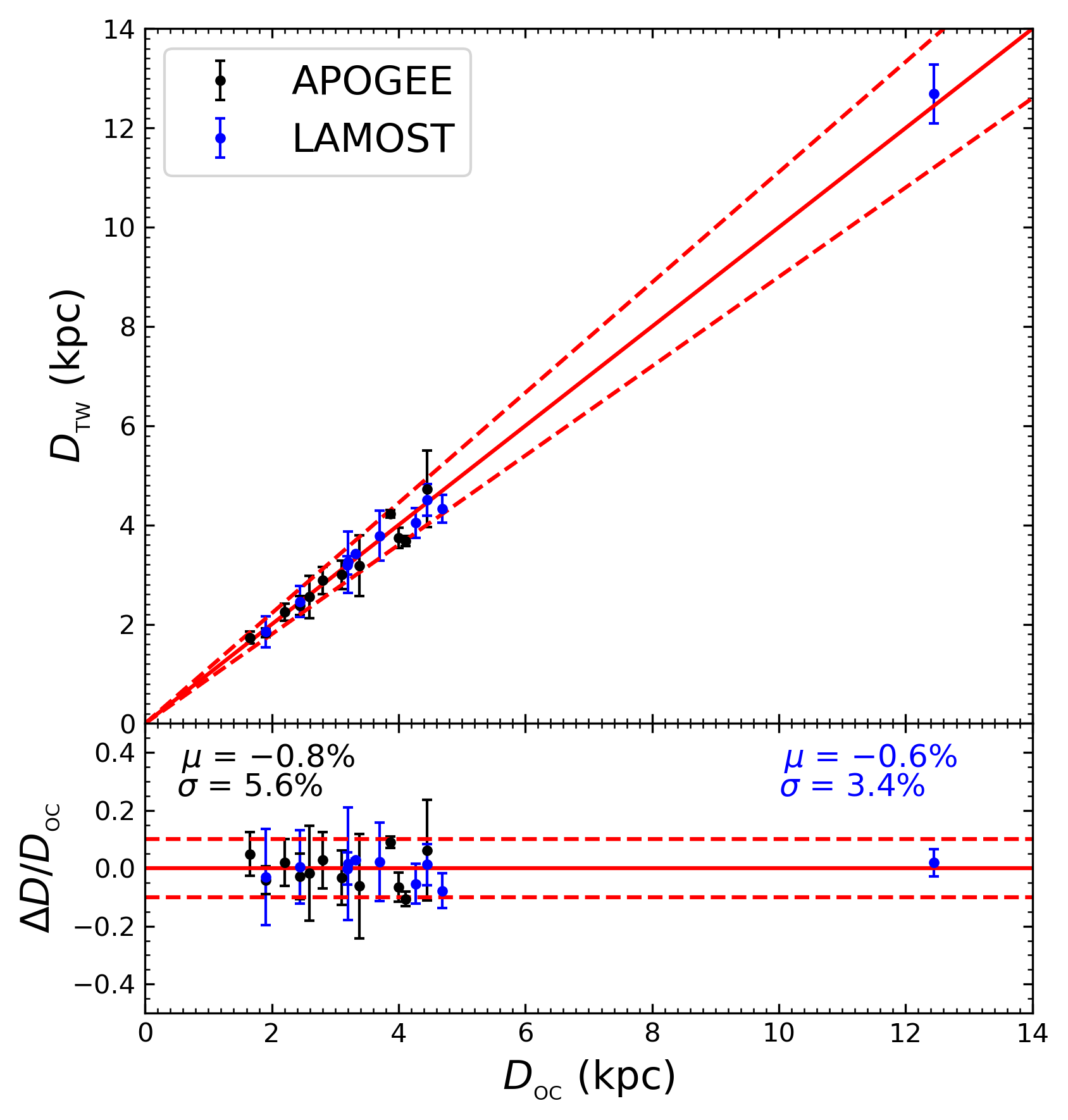}
\end{center}
\caption{Same as Fig.\,8 but for comparisons with OCs.}
\label{oc}
\end{figure}

In addition to the old metal-poor stars tested by GC members, we can perform similar checks for metal-rich young stars by using members of open clusters (OCs). \citet[][hereafter C18]{2018A&A...618A..93C} presented a catalog of about 45,000 member stars of 1,229 clusters based on the Gaia DR2 data by using UPMASK algorithm (Unsupervised Photometric Membership Assignment in Stellar Clusters) developed by \citet{2014A&A...561A..57K}. 
In total, 71/31 member stars of 12/10 OCs\footnote{The OCs used here are required to have number of members greater than two.} are found for APO-/LM-LRGB samples through cross-matching with the catalog of C18, with further clipping of few outliers by using line-of-sight velocity and metallicity estimated from the APOGEE and LAMOST. Comparisons of our distances of OCs with those published in the literature \citep[e.g.][]{2003MNRAS.340.1249P,2006MNRAS.368.1971B,2010AJ....140..954C,2014MNRAS.438.1451L,2020RMxAA..56..245A,2020arXiv201215587C,2021MNRAS.504..356D} are shown in Fig.\,\ref{oc} and Table\,3. 
Our spectrophotometric distances of OCs agree very well with those distances determined previously, with a negligible median offset of the relative distance difference $\Delta D/D$ of $-$0.8\%/$-$0.6\% and a standard deviation value of 5.6\%/3.4\% for APO/LM-LRGB samples. Finally,  the mean dispersions of the member distances revealed by individual OCs are all smaller than 10 cent, compared to the clusters' distances, both for APO-LRGB and LM-LRGB samples, again in agree with the results found for the training and testing samples. 
\begin{table*}
\centering
\begin{threeparttable}
\caption{Comparison of our distances of OCs with those published in the literature.}
\setlength{\tabcolsep}{1.5mm}{
\begin{tabular}{cccccccc}
\hline
\hline
Name & RA & DEC & RV& $D_{\rm{OC}}$ & $D_{\rm{TW}}$ &$N$ &Reference \\
& (degree) & (degree) & (km\,s$^{-1}$) & (kpc) & (kpc)&  \\
\hline
&  &  & (a)\,APO-LRGBs &  & &  \\
\hline
 Berkeley 17  & 80.126 & 30.574 & $-$72.502 &3.14$\pm$0.16 &3.00$\pm$0.29          &3  &D21\\ollinder 261&189.512&$-$68.378&$-$24.654  &2.81$\pm$0.12 &2.89$\pm$0.27            &3&D21\\
 IC 1848  &42.909&60.417&$-$57.452        &2.20$\pm$0.07    &2.24$\pm$0.18           &4&L14\\
 King 7   &59.770 &51.780    &$-$9.469       &2.59$\pm$0.36    &2.55$\pm$0.43       &6 & D21\\
 NGC 188  &11.824&85.241&$-$41.602       &1.65$\pm$0.05    &1.73$\pm$0.12           &2& C10\\
 NGC 2204  &93.882&$-$18.679&92.492      &4.11$\pm$0.12     &3.68$\pm$0.10         &9 &D21\\
 NGC 2243  &97.392&$-$31.287&59.611      &4.00$\pm$0.09    &3.74$\pm$0.20           &3& D21\\
 NGC 2304  &103.803&17.985&62.074        &3.87$\pm$0.15      &4.22$\pm$0.08         &2 & D21\\
 NGC 6791  &290.224&37.776&$-$47.258      &4.45$\pm$0.10    &4.72$\pm$0.77        &15 & D21\\
 NGC 6819  &295.329&40.190&3.376          &2.44$\pm$0.05    &2.38$\pm$0.19        &7 & D21\\
 NGC 7789  &342.296&56.727&$-$54.059     &1.91$\pm$0.03      &1.83$\pm$0.09       &9 &D21\\
 Trumpler 20 &189.849&$-$60.645&$-$38.785  &3.39$\pm$0.14   &3.18$\pm$0.61        &8  &D21\\
\hline
&  &  & (b)\,LM-LRGBs &  & &  \\
\hline

 Berkeley 29  &103.268&16.930& $-$        &12.45$\pm$1.41        &12.69$\pm$0.59   &4  & B06\\
 Berkeley 32  &104.528&6.430&106.417    &3.32$\pm$0.06        &3.42$\pm$0.01       &6 & D21\\
 Czernik 21  &81.675&36.011&45.770    &4.27$\pm$0.73        &4.04$\pm$0.30          &4 & D21\\
 NGC 1245  &48.685&47.236&$-$30.377      &3.19$\pm$0.11      &3.67$\pm$0.22      &3   & C20\\
 NGC 2158 &91.862&24.099 &26.900            &4.69$\pm$0.20    &4.33$\pm$0.28    &3  & A20\\
 NGC 2194  &93.451&12.813&41.263               &3.20            &3.25$\pm$0.62     &3  & P03\\
 NGC 2324  &106.029&1.046&43.948         &3.70$\pm$0.10     &3.78$\pm$0.50       &2 & D21\\
 NGC 6791  &290.224&37.776&$-$47.258      &4.45$\pm$0.10    &4.50$\pm$0.32        &2 & D21\\
 NGC 6819  &295.329&40.190&3.376         &2.44$\pm$0.05    &2.46$\pm$0.31          &2 & D21\\
 NGC 7789  &342.296&56.727&$-$54.059      &1.91$\pm$0.03    &1.85$\pm$0.32       &3  &D21\\
\hline
\end{tabular}}
\begin{tablenotes} 
\item Col.\,1 is name of the OC, Cols.\,2$-$5 give the central right ascension, declination, mean radial velocity and distances of the OCs from published literatures, Cols.\,6 presents the mean value of our distance and it error, and Col.\,7 gives the number of OC member stars from the (a)\,APO-LRGBs or (b)\,LM-LRGBs. The last column shows the references: \citet[][P03]{2003MNRAS.340.1249P}, \citet[][B06]{2006MNRAS.368.1971B}, \citet[][C10]{2010AJ....140..954C}, \citet[][L14]{2014MNRAS.438.1451L}, \citet[][A20]{2020RMxAA..56..245A}, \citet[][C20]{2020arXiv201215587C}, \citet[][D21]{2021MNRAS.504..356D}. 
\end{tablenotes} 
\end{threeparttable} 
\label{occ}
\end{table*}

To conclude, the extensive internal and external tests show that the spectrophotometric distances derived in this work are better than 10-15 per cent without significant systematics for the LRGB stars. The accurate distances will allow us to probe the properties of structures and kinematics of both disk(s) and halo of our Galaxy in a large volume, far beyond the limit (6-8\,kpc) of Gaia parallax.

\section{The Circular Velocity Curve}
\subsection{Kinematical model}
To derive the circular velocity curve, we only consider stars within $\vert\phi\vert\leqslant$ $\pm$ 30$\degr$ (the Sun–Galactic center line is set to $\phi = 0\degr$). 
To exclude the potential contaminations from stellar halo and thick disk, further cuts on positions: $|Z| \leqslant 0.5$\,kpc or $|b| \leqslant 6^{\circ}$, kinematics: $V_{Z} \leqslant 100$\,km\,s$^{-1}$ and chemistry: [Fe/H]\,$> -1.0$\,dex and [$\alpha$/Fe]\,$< 0.12$\,dex are applied. Eventually, the cuts result in 53,409 giant stars for the determination of RC. Assuming an axisymmetric gravitational potential $\Phi$ of the Milky Way, the rotation velocity $V_{\rm{c}}$($R$) of the Galaxy can be written as:

\begin{equation}
V_{\rm{c}}^2\left( R\right)=R \frac{\partial \Phi}{\partial R}.
\end{equation}

The Jeans equation \citep{2008gady.book.....B} in the cylindrical coordinate system could be expressed as, 
\begin{equation}
\boldsymbol{\frac{\partial (\nu \langle V_R^2 \rangle)}{\partial R}+\frac{\partial (\nu \langle V_R V_z \rangle)}{\partial z}+\nu \left( \frac{\langle V_R^2 \rangle -\langle V_\phi^2 \rangle}{R}+\frac{\partial \Phi}{\partial R}\right)=0,}\\
\end{equation}here $\nu$ is the density distribution of the tracer population. The cross-term $\langle V_R V_z \rangle$ and its vertical gradient is much smaller than the contributions from other terms, and is neglected in the following analysis. 
By simplifying Eq.\,(8) and combing with Eq.\,(7), the circular velocity can be derived by
\begin{equation}
V_{\rm{c}}^2\left( R\right)=\langle V_\phi^2 \rangle - \langle V_R^2 \rangle 
\left(1+ \frac{\partial \ln\nu }{\partial \ln R}+ \frac{\partial \ln\langle V_R^2 \rangle }{\partial \ln R} \right).
\end{equation}

In order to obtain the circular velocity $V_{\rm{c}}$, we need to know the radial density profile for the adopted tracer population. It is hard to precisely determine the scale length of the concerned tracers by our spectroscopic samples due to their poorly known selection functions. On the another hand, this scale length has been measured extensively by photometric samples and is well-known in the range of 2.0 to 2.6\,kpc \citep[e.g.,][]{2002ApJ...578..151S,2008ApJ...673..864J,2013ApJ...778...32P,2015PASA...32...12Y,2017MNRAS.464.2545C} and here we adopted a value of $2.13$\,kpc measured from red giant stars (similar to the tracers we adopted here) by \citet{2018MNRAS.477.2858W}.

\begin{figure*}
\begin{center}
\includegraphics[width=0.45\textwidth]{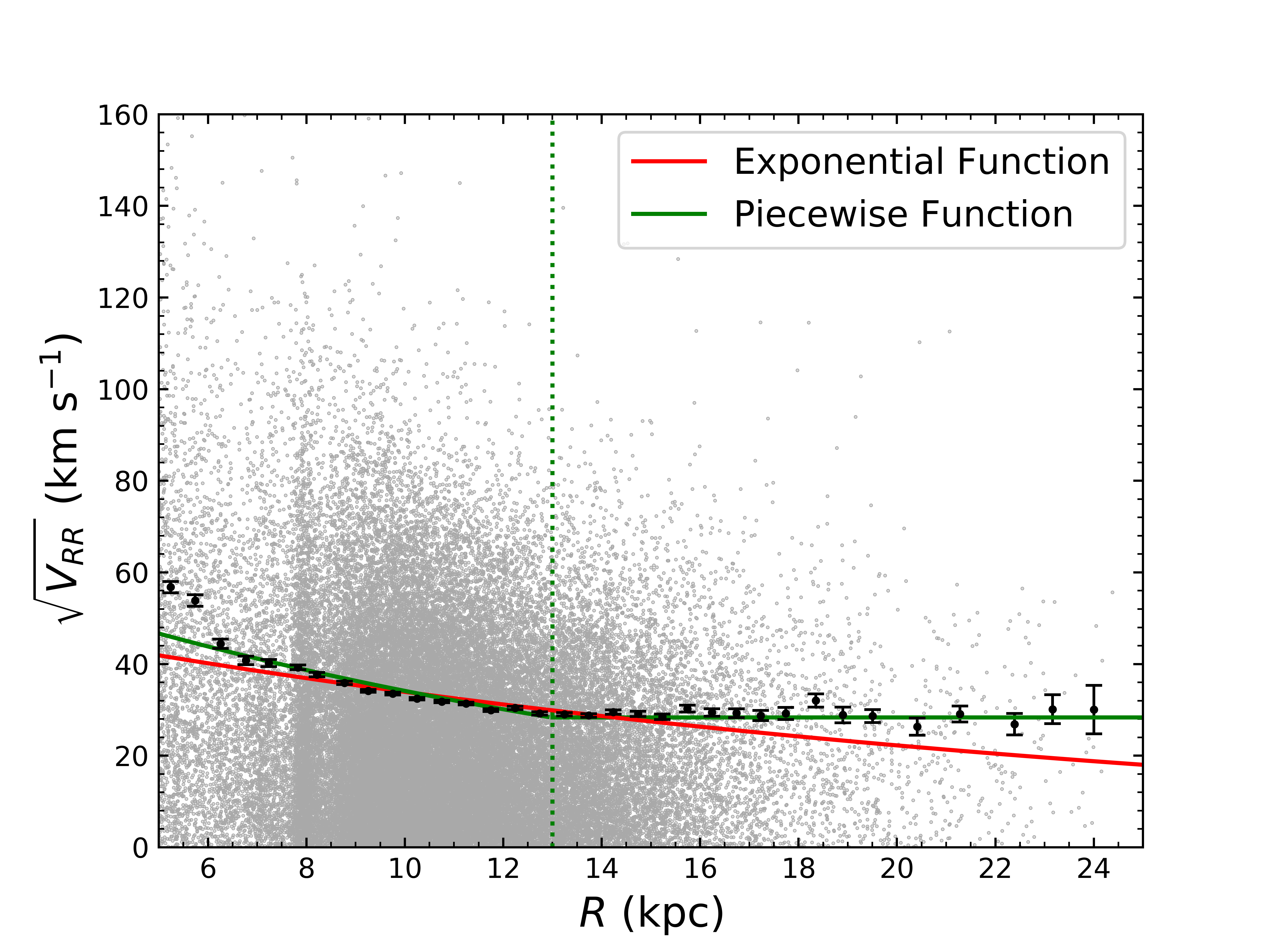}
\includegraphics[width=0.45\textwidth]{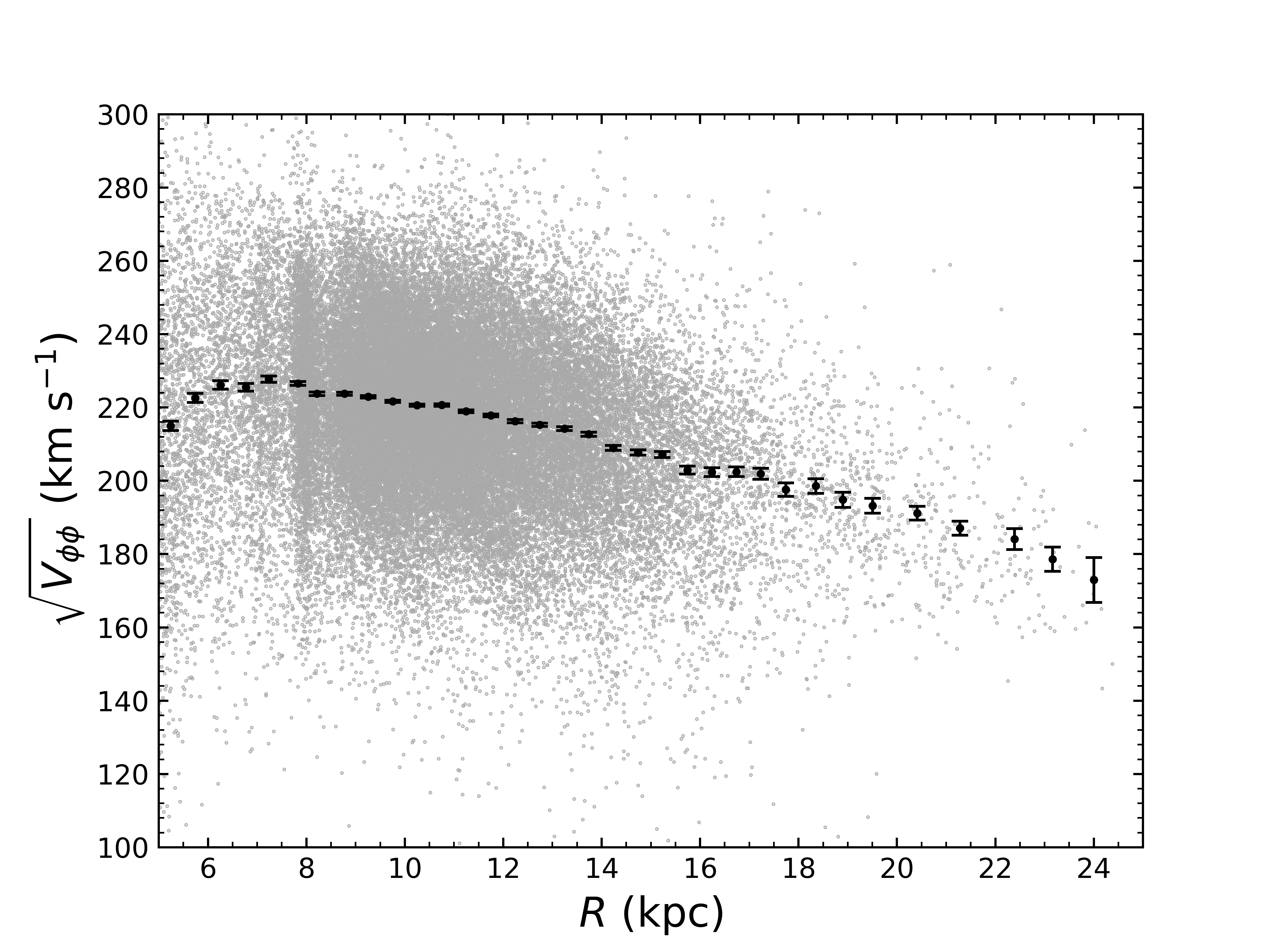}
\end{center}
\caption{Radial profiles of the components of the velocity tensor $\sqrt{V_{RR}}$ (left panel) and $\sqrt{V_{\phi\phi}}$ (right panel). The background grey dots show the results of individual stars, while black data dots represent the mean values of the velocity tensor in the individual radial annuli (the choice of binsize is described in Section 4.1), and their errorbars are evaluated by bootstrapping with 100 samples. The red and green lines in the left panel indicate the best-fits of $\sqrt{V_{RR}}$ by using an exponential function and a piecewise function, respectively (see text for details).}
\label{78}
\end{figure*}

The other terms in Eq.\,(9) can be measured directly from our sample. Fig.\,\ref{78} shows the radial profiles of the components of the velocity tensor $\sqrt{V_{RR}}$ (left panel), $\sqrt{V_{\phi\phi}}$ (right panel). $V_{RR}$ and $V_{\phi\phi}$ denote $V_R^2$ and $V_{\phi}^2$, and their means derived from an ensemble stars in a specific radial bin measure the values of $<V_R^2>$ and $<V_{\phi}^2>$ that are required to derive $V_{\rm c}$ according to Eq.\,(9). Doing so, we calculate the mean values of velocity tensors by dividing sample stars into different radial bins. The radial binsize is allowed to vary to contain at least 10 stars but no smaller than 0.5\,kpc. The results and their uncertainties yielded by bootstrapping are represented by black dots in Fig.\,\ref{nh}. Here we present two notes: 1) the uncertainties of $V_R$ are largely smaller than 5-10\,km\,s$^{-1}$ and thus can be ignored in deriving $<V_R^2>$, which is typically greater than (25\,km\,s$^{-1})^2$; 2) any stars within $R < 5$\,kpc are not considered in this study since the inner kinematics are very complicate to be modeled due to the presence of a central bar. 
Now the only unknown term in Eq.\,(9) to derive $V_{\rm c}$ is the scale length of radial velocity dispersion that also can be measured by performing an exponential fit to the measured radial profile of $\sqrt{V_{RR}}$. Our best-fit shows a scale length $R_{\sigma}$ of 24\,kpc for the radial velocity dispersion of the adopted tracer (see the red dotted line in Fig.\,\ref{78}), which is consistent with previous studies very well (e.g., \citealt{2016MNRAS.463.2623H}; E19). We note that the outer part of the profile of $\sqrt{V_{RR}}$ is not well described by a single exponential fit, due to the presence of well-known disk flaring \citep[e.g.][]{2002A&A...394..883L,2006A&A...451..515M,2014A&A...567A.106L,2021ApJ...912..106Y}. However, the discrepancies between the observational values of $\sqrt{V_{RR}}$ and the fitted ones are negligible on deriving RC (see discussions in section 4.2.2).

\subsection{Results}
\subsubsection{The circular velocity curve}
\begin{figure*}
\begin{center}
\includegraphics[width=0.925\textwidth]{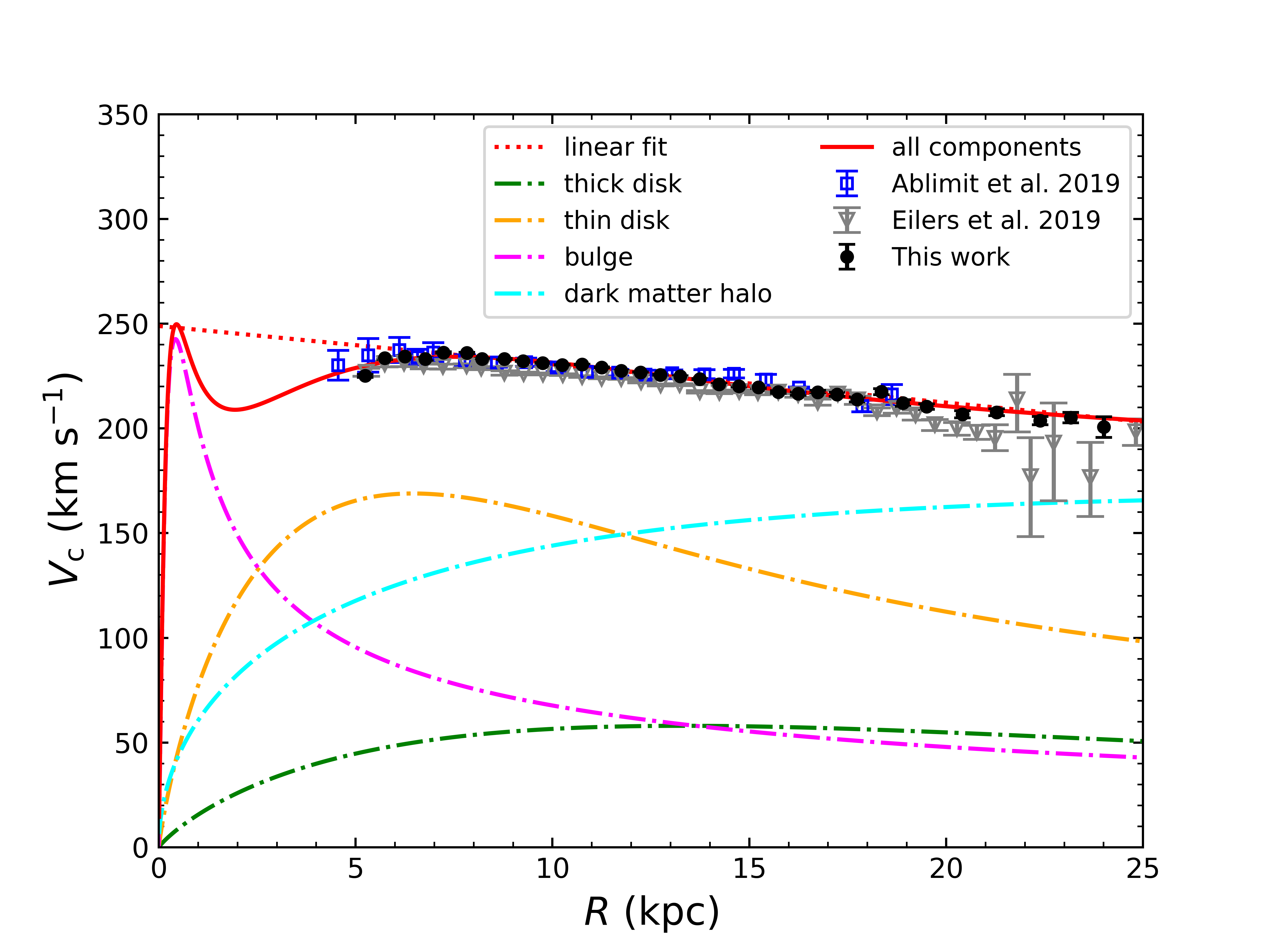}

\end{center}
\caption{New measurements of the RC of the Milky Way are shown as the black circles with the errorbars estimated by bootstrapping (systematics not included here). The recent measurements from \citet{2019ApJ...871..120E} and \citet{2020ApJ...895L..12A} are shown as colored symbols for comparisons. The red dotted line shows a linear fit to our RC. The red solid line represents the best-fitted model RC to the data points, which is the sum of the contributions of the bulge (magenta dot-dashed line), the thin disk (golden dot-dashed line), the thick disk (green dot-dashed line), and the dark matter halo (cyan dot-dashed line).}
\label{nh}
\end{figure*}

Now we can calculate circular curves $V_{\rm c}$  of the Milky Way via Eq. (9). The results are shown in Fig.\,\ref{nh} and presented in Table\,\ref{tfrc}.  This curve is the most accurate one to date with uncertainties smaller than 1.0\,km\,s$^{-1}$ for $R < 18$\,kpc and 1.0-3.0\,km\,s$^{-1}$ for $R > 18.0$\,kpc. Compared to E19, the precision of our RC is 2-3 times better for $R$ within 18\,kpc and 5-10\,times improved beyond that radius. For inner part ($R < 18$\,kpc), the newly derived RC is generally consistent with that of E19 very well. However, the most outer part ($R > 18$\,kpc) of the E19's RC drops too fast and can not be reproduced by the canonical mass models (see Fig.\,3 of E19), while our newly derived RC is much improved in this region (see Fig.\,\ref{nh}). The overall RC shows a gently decreasing trend with $R$, as found by many recent studies. By a linear fit shown in the Fig.\,\ref{nh}, a negative slope of $-1.83 \pm 0.02$\,km\,s$^{-1}$\,kpc$^{-1}$ is found, in great consistent with the slope of $-1.7 \pm 0.1$ \,km\,s$^{-1}$\,kpc$^{-1}$ derived by E19, and slightly higher than the gradients of $-1.33$ to $-1.44$\,km\,s$^{-1}$\,kpc$^{-1}$ found by \citet{2019ApJ...870L..10M} and \citet{2020ApJ...895L..12A} in the radial region of $4 \leqslant R \leqslant 18$\,kpc using hundreds to thousands of classical Cepheids. Moreover, the linear fit yields a local circular speed $V_{\rm c} (R_0)=234.04 \pm 0.08$\,km\,s$^{-1}$, which is consistent with recent determinations very well \citep[e.g.,][]{2019ApJ...871..120E,2019ApJ...870L..10M,2020ApJ...895L..12A,2020MNRAS.494.6001N}. 

\begin{table}
\centering
\begin{threeparttable}
\caption{Measured values and errors of the circular velocity of the Milky Way.}
\label{tfrc}
\setlength{\tabcolsep}{1.5mm}{
\begin{tabular}{cccc}
\hline
\hline
$R$ (kpc) & $V_{\rm{c}}$ (km\,s$^{-1}$) & $\sigma_{V_{\rm{c}}}$ (km\,s$^{-1}$) & Number\\
\hline
 5.24&225.10&0.69&845\\
 5.74&233.53&0.68&692\\
 6.25&234.3&0.62&704\\
 6.77&233.17&0.60&759\\
 7.23&236.19&0.45&1061\\
 7.83&236.00&0.29&2288\\
 8.21&233.19&0.26&2550\\
 8.78&233.15&0.22&3281\\
 9.26&232.15&0.17&4583\\
 9.75&231.24&0.16&5061\\
 10.25&230.34&0.17&4881\\
 10.75&230.54&0.18&4564\\
 11.25&229.11&0.19&4005\\
 11.75&227.48&0.20&3431\\
 12.24&226.69&0.25&2844\\
 12.74&225.56&0.27&2312\\
 13.25&224.90&0.27&2116\\
 13.74&223.57&0.31&1825\\
 14.23&221.10&0.40&1362\\
 14.74&220.19&0.43&987\\
 15.23&219.59&0.50&801\\
 15.74&217.36&0.68&563\\
 16.24&216.61&0.74&446\\
 16.74&217.28&0.87&308\\
 17.23&216.25&1.02&257\\
 17.74&213.81&1.15&163\\
 18.35&217.53&1.45&207\\
 18.90&212.10&1.58&97\\
 19.50&210.46&1.32&162\\
 20.41&206.69&1.71&85\\
 21.28&207.71&1.69&93\\
 22.39&203.72&2.01&46\\
 23.16&205.20&2.50&20\\
 24.00&200.64&4.94&10\\
 
\hline
\end{tabular}}
\begin{tablenotes} 
\item Columns show the median Galactocentric radius, the circular velocity, the errorbars and the number of star for each radial bin.
\end{tablenotes} 
\end{threeparttable} 
\end{table}

\subsubsection{Systematic uncertainties}

\begin{figure}
\begin{center}
\includegraphics[width=0.5\textwidth]{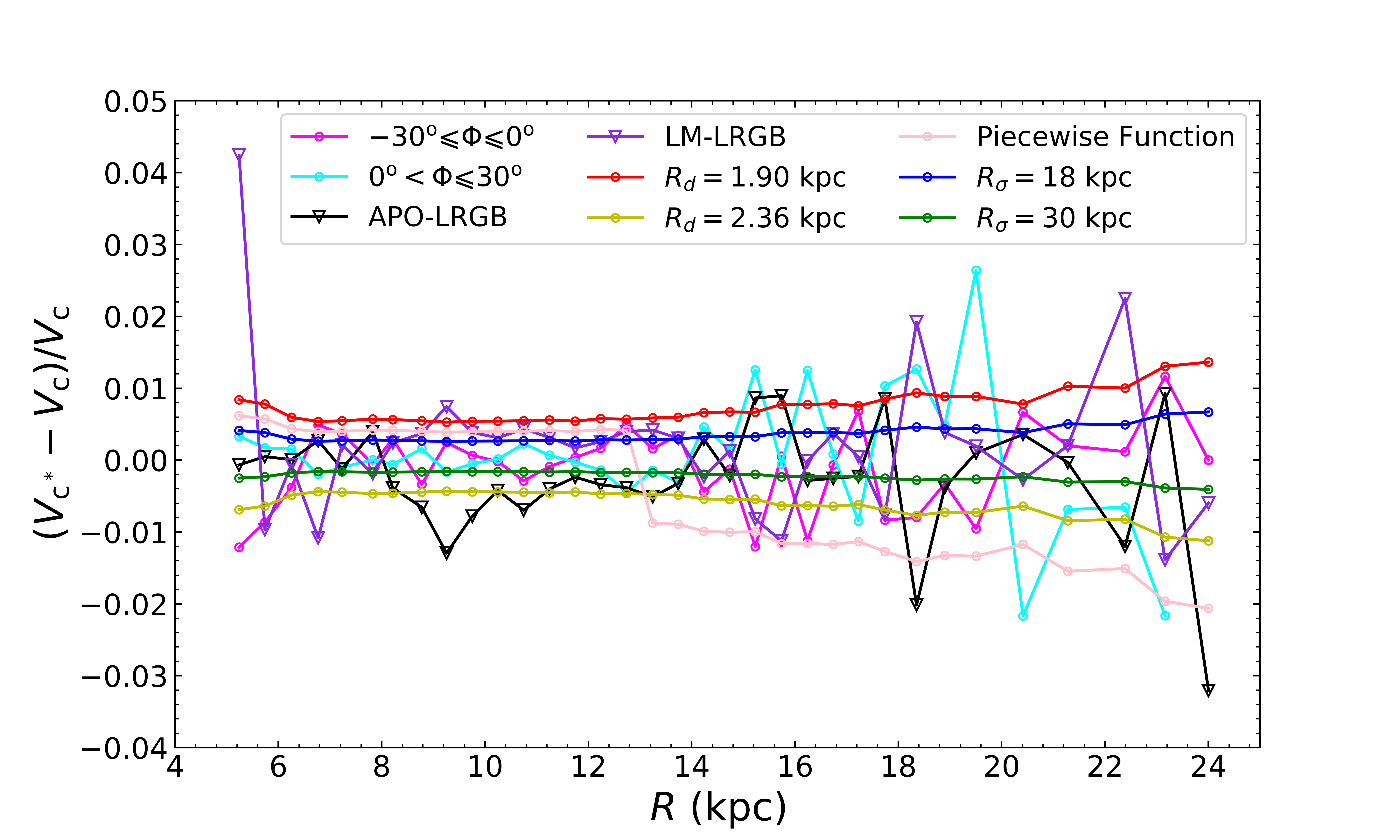}
\end{center}
\caption{Potential systematic uncertainties, $(V_{\rm c}^* - V_{\rm c})/V_{\rm c}$, as a function of $R$. 
$V_{\rm c}$ represents the circular speed deduced in Section\,4.2.1 while $V_{\rm c}^*$ indicates the circular speed derived with alternative parameters or sub-samples as marked in the right-top corner.}
\label{xt}    
\end{figure}

In this section we discuss the potential systematic  effects on the measurements of the RC. 
The evaluated systematic uncertainties as a function of the Galactocentric radius, $(V_{\rm c}^* - V_{\rm c})/V_{\rm c}$, from various contributions are summarized in Fig.\,\ref{xt}. Here $V_{\rm c}$ and $V_{\rm c}^*$ represent the circular velocity derived in Section\,4.2.1 and the velocity derived by considering different systematic effects, respectively.

Our thin disk star sample is obtained by merging two LRGB samples, built from the LAMOST optical spectra and APOGEE near-infrared spectra, respectively. We then re-calculate the RCs from APO-LRGB and LM-LRGB samples, to examine any differences of the established RCs by the two samples.The results are shown in Fig.\,\ref{xt} and one can clearly find that the systematic effects are largely smaller than 1.0 per cent for both APO-LRGB (black downward-pointing triangles) and LM-LRGB (purple downward-pointing triangles) in all radial bins.

Another systematic uncertainty due to the data sample could be the differences arising from sample stars of different azimuthal ranges. 
To assess this systematic uncertainty, we divide the sample stars into two sub-samples with $-30\degr \leqslant \phi \leqslant 0\degr$ and $0\degr \leqslant \phi \leqslant 30\degr$, respectively, and performed the same analysis for the two sub-samples. At the location of the Sun we determine a relative difference of 0.29\% for the region with $-30\degr \leqslant \phi \leqslant 0\degr$ and of $-$0.06\% for the region with $0\degr<\phi \leqslant 30\degr$. It is not difficult to find from the magenta and cyan open circles in Fig.\,\ref{xt} that the systematic uncertainties do not exceed $\sim$1\% overall.

As mentioned in E19, the largest systematic uncertainties on the measured circular velocities at most radial bins are the unknown scale length of the adopted tracer population. In Section\,4.2.1, we adopt an exponential function with a commonly used scale length $R_d =$ 2.13\,kpc to describe the density profile of our LRGB stars. To evaluate this systematic uncertainty, we try two alternative values, i.e. a lower value of 1.90\,kpc and a higher value of 2.36\,kpc to repeat the whole analysis. At the location of the Sun, a relative difference of $0.57$\% and $-0.46$\% is found for $R_d =1.90$\,kpc and $R_d = 2.36$\,kpc, respectively. As shown in Fig.\,\ref{xt}, this systematics increases with $R$ but is still smaller than 1.0-1.5 per cant at $R = 25$\,kpc. 

As mentioned earlier and also seen in the left panel of Fig.\,\ref{78} (red line), the profile of $\sqrt{V_{RR}}$ can not be well fitted by a single exponential function in the outer disk with $R > 14$\,kpc, where the fitted values are significantly lower than the observational ones. To better describe this profile, we repeat the fit by a piecewise function, with a broken radius at $R_b = 13$\,kpc. Within $R_b$, an exponential function with a scale length of $16$\,kpc is found to agree the data points very well while a straight line with a constant value of $28.4$\,km\,$^{-1}$ is consistent with the observations beyond $R_b$ (green line in the left panel of Fig.\,\ref{78}). However, an unphysical drop around $R_b$ will be induced in deriving RC if adopting this piecewise function rather than the single exponential. Meanwhile, as examined in Fig.\,\ref{xt}, the systematic effects, due to the unsuccessful fits to the outer part of the profile of $\sqrt{V_{RR}}$ by a single exponential function, are very minor, in the level of 1.0\%. A single exponential function is therefore a not bad choice on describing the profile of radial velocity dispersion for the purpose of deriving RC.

We also test two extreme values of $R_{\sigma}$, i.e., 18 and 30\,kpc, and the effects on RC measurements are minor, within the level of 1 per cent in all radial bins (see blue and dark green circles on Fig.\,\ref{xt}). 
As mentioned earlier, the cross-term is ignored on deriving RC, that causes a systematic uncertainty typically smaller than 1 per cent for $R < 18$\,kpc and slightly larger beyond this radius (see the discussion of E19).

Finally, we also explore all the mentioned effects on determining the RC gradient  and the local circular speed and find their systematic uncertainties are about 0.07\,km\,s$^{-1}$\,kpc$^{-1}$ and 1.36\,km\,s$^{-1}$.

\section{Galactic mass models based on the newly RC} 
Although the field of Galactic dynamics is reaching an incredibly exciting time, the distribution of mass in the various components, especially the dark matter halo, of the Milky Way is still poorly determined (see the recent review of \citealt{2020SCPMA..6309801W}). As mentioned earlier, the RC is a power tool to measure the mass distribution of our Galaxy. The newly RC constructed here is therefore used to constrain the Galactic mass distribution. Doing so, a parametrized mass model of the Milky Way is constructed to fit the observed RC.
In this model, four major components, i.e., the central bulge, the thin and thick disks, and the dark matter halo, are considered. The model predicted circular velocities as a function of Galactic radius can be calculated from the contributions of the four components,

\begin{equation}
V_{\rm{c}}^2=V_{\rm{c,bugle}}^2+V_{\rm{c,thin{\,}disk}}^2+V_{\rm{c,thick{\,}disk}}^2+V_{\rm{c,halo}}^2.
\end{equation}
We briefly describe the individual components as following.

(1) The bulge. 
Since we have no data about RC within 5\,kpc to restrict the bulge mass distribution, here we fix the parameters of the bulge, i.e. using the Plummer bulge mass model constructed by \citet{2011MNRAS.413.1889B}. The density distribution of the bulge is simply described by

\begin{equation}
\rho_{\rm{bugle}}=\frac{3b^2M_b}{4\pi(r^2+b^2)^{5/2}}.
\end{equation}

The contribution to circular velocity of the bulge could be given by,
\begin{equation}
V_{\rm{c,bugle}}^2=\frac{GM_br^2}{(r^2+b^2)^{3/2}},
\end{equation}
here, $M_{\rm{b}} = 1.067 \times$10$^{10} \rm{M_\odot}$ and $b =$ 0.3\,kpc are the bulge mass and scale constants, directly adopted from \citet{2011MNRAS.413.1889B}.

(2) The disks. 
The Milky Way’s stellar disk is commonly decomposed into two sub-components, a thin and a thick disk \citep{1983MNRAS.202.1025G}. The Galactic disk is represented by an exponential disk \citep{1970ApJ...160..811F} and its surface density is described by,
\begin{equation}
\Sigma\left( R\right) =\Sigma_{\rm{d,0}}{\rm{exp}}\left( -\frac{R}{R_{\rm{d}}}\right). 
\end{equation}\\with a central surface density $\Sigma_{\rm{d,0}}$ and a scale length $R_{\rm{d}}$. The total mass is $M_{\rm{d}} = 2\pi\Sigma_{\rm{d,0}} R_{\rm{d}}^2$. In this paper, we regard the disk as the exponential razor-thin disk, and the circular velocity is therefore given by,
\begin{equation}
V_{\rm{c,disk}}^2=4\pi G \Sigma_{\rm{d,0}}R_{\rm{d}}y^2[I_0(y)K_0(y)-I_1(y)K_1(y)],
\end{equation}

where $y = R/(2R_{\rm{d}})$, $I_n$ and $K_n$ (n = 0, 1) are the first and second kind modified Bessel functions, respectively \citep{1987gady.book.....B}. 
Here, the surface densities of both thin and thick disks are fixed to the local measurements: $\Sigma_{{R_{\rm{0}},\rm{thick}}} =7.0$ $\rm{M_\odot}{\rm{pc}}^{-2}$ \citep{2006MNRAS.372.1149F} and $\Sigma_{{R_{\rm{0}},\rm{thin}}} =30.4$ $\rm{M_\odot}{\rm{pc}}^{-2}$ \citep{2016MNRAS.463.2623H}. 
The central surface density is then easily derived by $\Sigma_{\rm{d,0}} =\Sigma_{R_{\rm{0}}}\rm{exp}(R_{\rm 0}/R_{\rm d})$.

(3) The dark matter (DM) halo. We considered the canonical NFW profile \citep{1995MNRAS.275...56N} to describe the DM halos in spherical coordinates:

\begin{equation}
\rho_{\rm{halo}}(r)=\frac{\rho_{\rm{0}}}{(r/r_{\rm{s}})(1+r/r_{\rm{s}})^2},
\end{equation}
where $r_{\rm{s}}$ is the scale radius, and $\rho_{\rm{0}}$ is the characteristic DM density. Here we adopt the definition of the total mass, i.e. $M_{200},$ of the halo within a spherical volume of the radius, i.e. $R_{200}$, with mean density 200 times of the critical density, $\rho_{\rm crit}$ of the universe.
By adopting the cosmological parameters from \citet{2013ApJS..208...19H}, this density is $1.34 \times 10^{-7} M_{\odot}$\,pc$^{-3}$. The circular speed from NFW DM is then calculated by:

\begin{equation}
V_{\rm{c,halo}}^2=\frac{4G\pi\rho_{\rm{0}}r_{\rm{s}}^3}{r}\left[{\rm{ln}}(1+\frac{r}{r_{\rm{s}}})-\frac{r}{r_{\rm{s}}(1+\frac{r}{r_{\rm{s}}})}\right]. 
\end{equation}

In total, there are four free parameters in our Galactic mass model: two for the disks ($R_{\rm{d,thick}}$, $R_{\rm{d,thin}}$) and two for the DM halo ($r_{\rm{s}}$, $\rho_{\rm{s}}$). Then, we fit the model circular velocities given by Eq.\,(10) to the observed ones to constrain the four free parameters. To do so, we apply the Markov Chain Monte Carlo (MCMC) affine invariant sampler emcee \citep{2013ascl.soft03002F} to sample the parameter spaces of the four free parameters by maximizing the likelihood defined as follows:

\begin{equation}
{\rm{ln}}{\ }p(V_{\rm{c}}|R,\theta)=-\frac{1}{2}\sum_{i=1}^N[\frac{(V_{{\rm c},i}^{\rm{obs}}-V_{{\rm c},i}^{\rm{mod}})^2}{\sigma_{i}^2}+{\rm{ln}}(2\pi\sigma_{i}^2)],
\end{equation}where $\theta$ represents the four free parameters mentioned above. $V_{{\rm c},i}^{\rm{obs}}$ represents the model predicted circular velocity, $V_{{\rm c},i}^{\rm{mod}}$ represents the observed circular velocity, $\sigma$ represents the statistical uncertainty of the measurement and $N$ represents the number of radial bin.

\begin{figure*}
\begin{center}
\includegraphics[width=0.9\textwidth]{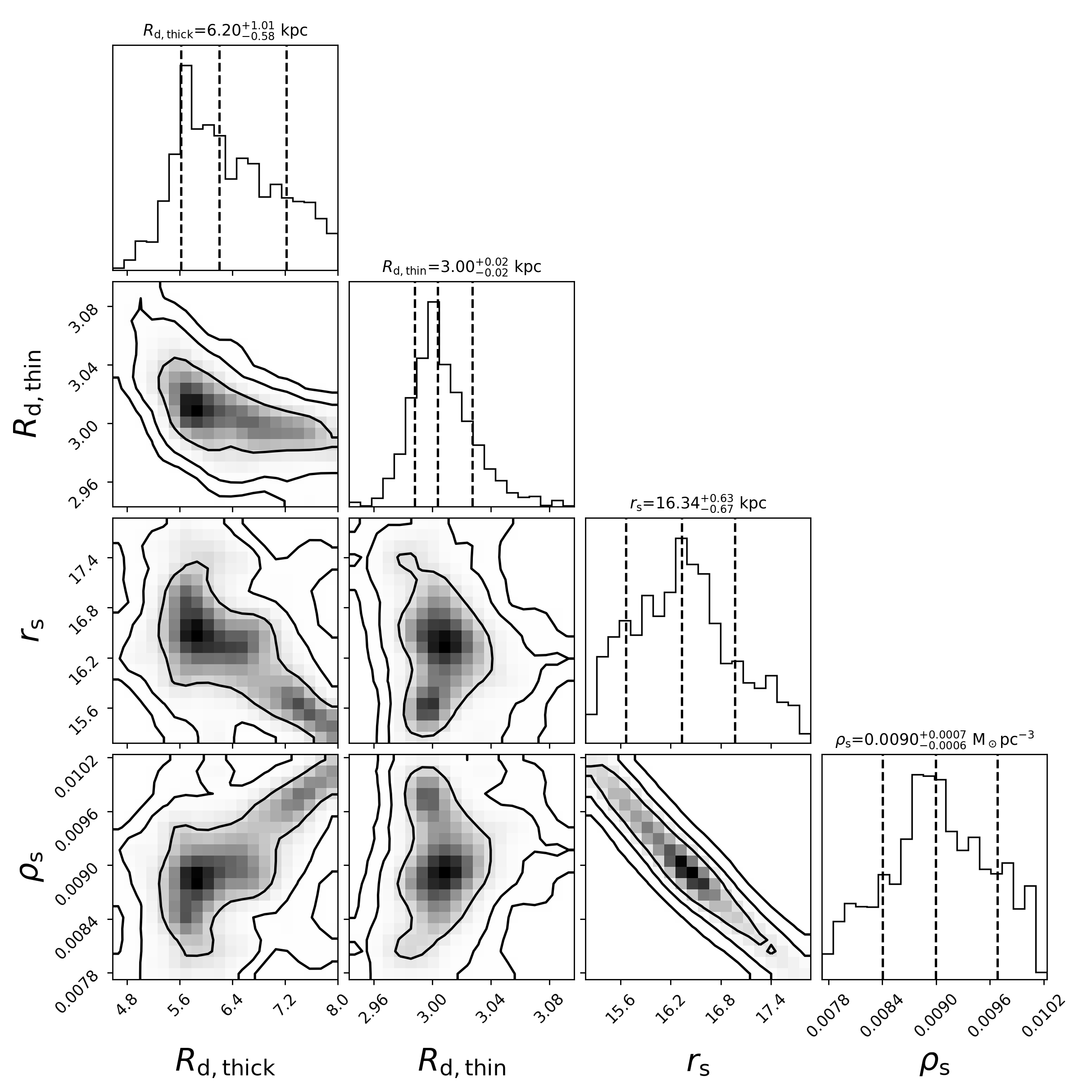}
\end{center}
\caption{One- and two-dimensional marginalized PDFs of four model parameters obtained from the MCMC. 
Contours in each panel mark the $1\sigma$,  $2\sigma$ and  $3\sigma$ regions, respectively.
The histograms at the top of each column show the one-dimensional marginalized PDFs of each parameter. 
The dashed lines in each top histogram represent the 16 per cent, 50 per cent, 84 per cent percentiles of the distribution, respectively. 
The best-fitted values and their uncertainties of the model parameters are marked at the top of each column.}
\label{123123}    
\end{figure*}

Fig.\,\ref{123123} shows the posterior probability distribution functions (PDFs) of the four free parameters given by the MCMC. The best values, as well as the uncertainties, of the four parameters are estimated by the 50\%, and the 16\% and 84\% percentiles of the resultant posterior PDFs. 
As shown in Fig.\,\ref{nh}, the model predicted RC can describe the observed one very well for the data points in all radial bins.

In the best-fitted mass model, the mass of the DM of the Milky Way is $M_{\rm{200}}$ = (8.05 $\pm$ 1.15) $\times$10$^{11} \rm{M_\odot}$ and the concentration parameter is $c$ = 11.80 $\pm$ 0.73, with a corresponding radius $R_{\rm{200}} =$ 192.37 $\pm$ 9.24\,kpc. From $\rho_{\rm{s}}$ and $r_{\rm{s}}$, we find the local DM density is $\rho_{\odot}$ = 0.39 $\pm$ 0.03\,GeV\,cm$^{-3}$.

\begin{table*}
\centering
\begin{threeparttable}
\caption{Best-fitted mass model parameters.}
\label{parameter}
\setlength{\tabcolsep}{8.5mm}{
\begin{tabular}{ccccccccc}
\hline
\hline
Galactic component & Parameter & Value & Unit & Note\tnote{a}\\
\hline
Bulge
& $M_{\rm{b}}$ & 1.067 & 10$^{10} \rm{M_\odot}$&fixed\\
& $b$ & 0.3 & kpc&fixed\\
\hline
Thin disk  
& $\Sigma_{\rm{d,thin}}$ & 456.17$\pm$8.00 & $\rm{M_\odot}{\rm{pc}}^{-2}$&fixed\\
& $R_{\rm{d,thin}}$ & 3.00$_{-0.02}^{+0.02}$ & kpc&fitted\\
& $M_{\rm{d,thin}}$ & 2.57 $\pm$ 0.01 & 10$^{10} \rm{M_\odot}$&derived\\
\hline
Thick disk
& $\Sigma_{\rm{d,thick}}$ &26.63$\pm$ 4.78 & $\rm{M_\odot}{\rm{pc}}^{-2}$&fixed\\
& $R_{\rm{d,thick}}$ & 6.20$_{-0.58}^{+1.01}$ & kpc &fitted\\
& $M_{\rm{d,thick}}$  & 0.63 $\pm$ 0.05 & 10$^{10} \rm{M_\odot}$&derived\\
\hline
Halo  
& $r_{\rm{s}}$ & 16.34$_{-0.67}^{+0.63}$ & kpc&fitted\\
& $\rho_{\rm{s}}$  & 0.0090$_{-0.0006}^{+0.0007}$ & $\rm{M_\odot}{\rm{pc}}^{-3}$&fitted\\
& $\rho_{\rm{\odot}}$ & 0.39 $\pm$ 0.03 & GeV\,cm$^{-3}$&derived\\
& $c$ & 11.80 $\pm$ 0.73 & $-$&derived\\
& $M_{\rm{200}}$ & 8.05 $\pm$ 1.15 & 10$^{11} \rm{M_\odot}$&derived\\
& $R_{\rm{200}}$ & 192.37 $\pm$ 9.24 & kpc&derived\\
\hline
All
& $M_{\rm{total}}$ & 8.47$\pm$ 1.15& 10$^{11} \rm{M_\odot}$&derived\\
\hline
\end{tabular}}
\begin{tablenotes} 
\item $^{\rm a}$ Here, ‘fixed’, ‘fitted’ and ‘derived’ denote the concern parameter/quantity is either fixed or fitted in the Galactic mass model, or derived from the best-fitted model.
\end{tablenotes} 
\end{threeparttable}
\end{table*}

\section{Summary}
Based on massive spectra collected by the APOGEE and LAMOST surveys, combined with $J$ and $K_{\rm s}$ band photometric data from 2MASS and astrometric measurements from the Gaia surveys, we present a large LRGB sample of 254,882 stars with precise spectrophotometric distances estimated by the GBDT technology. 
Various tests show that the accuracy of estimated distance is better than 10-15\%. 
Stellar atmospheric parameters, line-of-sight velocities, elemental abundances and proper motions from the APOGEE, LAMOST and Gaia surveys are also provided for the sample stars.

By carefully selection of about 54,000 thin disk stars from our LRGB sample, the most accurate RC with $R$ from 5\,kpc to 25\,kpc is built via the Jeans model by assuming assuming an axisymmetric gravitational potential.
The RC shows a weak decline along $R$, as found by several recent studies.
A simple linear fit of this RC yields a slope of $-$1.83 $\pm$ 0.02 (stat.) $\pm$ 0.07 (sys.) \,km\,s$^{-1}$\,kpc$^{-1}$ that is similar to the results of recent determinations.
A local circular speed of $V_{\rm c} (R_0)$ = 234.04 $\pm$ 0.08 (stat.) $\pm$ 1.36 (sys.)\,km\,s$^{-1}$ is found again by this linear fit and the result is consistent with other recent independent measurements.

Based on the newly constructed RC, as well as other constraints on stellar components, a parametrized mass model is built for our Galaxy, yielding a total mass of the DM halo $M_{\rm{200}}$ = (8.05 $\pm$ 1.15) $\times$ 10$^{11}\rm{M_\odot}$ with a corresponding radius is $R_{\rm{200}}$ = 192.37 $\pm$ 9.24\,kpc.
The derived mass is in excellent agreement with the results of $(0.82 \pm 0.05) \times 10^{12} \rm{M_\odot}$ derived by \citet{2020ApJ...895L..12A}, 0.83$_{-0.09}^{+0.12} \times10^{12} \rm{M_\odot}$ derived by \citet{2020JCAP...05..033K}, 0.90$_{-0.08}^{+0.07} \times10^{12} \rm{M_\odot}$ derived by \citet{2016MNRAS.463.2623H} and lightly lager than the result of (7.25 $\pm$ 0.26) $\times 10^{11} \rm{M_\odot}$ derived by E19. 
The large uncertainty found in this study is mainly because we do not fix the contributions from the stellar disks in building Galactic mass model.
The best-fitted model also yields a local dark matter density $\rho_{\odot}$ of 0.39 $\pm$ 0.03\,GeV\,cm$^{-3}$ that is very close to the recent measurements of $0.30 \pm 0.11$\,GeV\,cm$^{-3}$ by \citet{2012ApJ...756...89B}, $0.32 \pm 0.02$\,GeV\,cm$^{-3}$ by \citet{2016MNRAS.463.2623H} and $0.33 \pm 0.03$\,GeV\,cm$^{-3}$ by \citet{2020ApJ...895L..12A}.

\section*{Acknowledgements} 
We thank an anonymous referee for helpful comments.
YH acknowledges the NSFC with No. of 11903027 and 11833006 and the National Key RD Program of China No. 2019YFA0405503. 
HWZ acknowledges the NSFC with No. of 11973001, 12090040, 12090044, and the National Key RD Program of China No.2019YFA0405504.
Xinyi Li is supported by the Yunnan University grant 2021Z015. 

This work presents results from the European Space Agency (ESA) space mission Gaia. Gaia data are being processed by the Gaia Data Processing and Analysis Consortium (DPAC). Funding for the DPAC is provided by national institutions, in particular the institutions participating in the Gaia MultiLateral Agreement (MLA). The Gaia mission website is {\url{https://www.cosmos.esa.int/gaia}}. The Gaia archive website is {\url{https://archives.esac.esa.int/gaia}}.

The Guoshoujing Telescope (the Large Sky Area Multi-Object Fiber Spectroscopic Telescope, LAMOST) is a National Major Scientific Project built by the Chinese Academy of Sciences. Funding for the project has been provided by the National Development and Reform Commission. LAMOST is operated and managed by the National Astronomical Observatories, Chinese Academy of Sciences. The LAMOST website is {\url{https://www.lamost.org}}.

Funding for the Sloan Digital Sky Survey IV has been provided by the Alfred P. Sloan Foundation, the U.S. Department of Energy Office of Science, and the Participating Institutions. SDSS-IV acknowledges
support and resources from the Center for High-Performance Computing at the University of Utah. The SDSS website is {\url{https://www.sdss.org}}.

\bibliographystyle{apj}
\bibliography{GRC}

\begin{appendix}

\section{Extinction from the ``star pair" method}
To improve the accuracy of the spectrophotometric distance by our method, it is necessary to estimate the values of extinction of individual stars precisely. Most of the program stars in this study are in the Galactic disk region with quite large values of extinction that are hardly corrected by the existing and widely adopted SFD map \citep{1998ApJ...500..525S}. This map is a whole-sky 2D dust-reddening map derived from thermal emission that is not suitable for disk area and very high-extinction regions \citep[e.g.][]{1999APS..DFD..JH01A,2010ApJ...725.1175S,2013MNRAS.430.2188Y}. Thanks to the precise estimates of atmospheric parameters from the LAMOST and APOGEE surveys, we can estimate precise extinction for every program stars by using the ``star pair" method \citep[e.g.][]{1965ApJ...142.1683S,2013MNRAS.430.2188Y}. 

Doing so, we first define a library of LRGB stars at high Galactic regions with very low values of extinction that their intrinsic colors can be precisely derived. To select such stars, we have adopted the following criteria: 1) spectral SNRs $> 50$; 2) $E (B-V)_{\rm SFD} < 0.05$ (corrected for a 14\% overestimated systematics); 3) Galactic latitude higher than 25 degree. In total, 14,268 and 14,825 stars are selected from APO-LRGBs and LM-LRGBs as the library. For each program stars, we then select the library stars with values of $T_{\rm eff}$ and [Fe/H] that differ the parameters of the program stars no more than 25\,K and 0.05\,dex. A minimum number of 50 library stars is required for each target and the bin size cuts could be slightly relaxed to meet the minimum number requirement. The intrinsic color $(J - K_{\rm s})_0$ of this program star is predicted by the linear relation between $(J - K_{\rm s})_0$ and stellar parameters (i.e., $T_{\rm eff}$ and [Fe/H] here) fitted for selected library stars. Finally, the colour excess $E (J - K_{\rm s})$ is derived by comparing the observed colour and predicted intrinsic color.

\section{Training and testing samples}
As mentioned in Section 3.1, it is of vital importance to select a training sample with precise absolute magnitudes and wide coverages in parameter spaces. 
The main part of the training sample (including the testing sample) is stars with precise absolute magnitude measured from Gaia parallax at low extinction regions that satisfy the following criteria:
\\$\bullet$ spectral SNRs\,$> 100$ (loosed to SNRs\,$> 50$ for distant stars with $d > 8.5$\,kpc);
\\$\bullet$  $M_{K_{\rm s}} < -1.0$\,mag, $e_{K_{\rm s}} < 0.030$\,mag and $e_{M_{K_{\rm s}}} <0.2$\,mag  (the latter two are loosed to 0.035 and 0.35\,mag, resepctively, for distant stars with $d > 8.5$\,kpc);
\\$\bullet$  $E(B-V) < 0.08$\,mag and $|b|> 25^{\circ}$ ($E(B-V)$ is loosed to smaller than 0.10\,mag for distant stars with $d > 8.5$\,kpc).

In total, 7,642 and 6,798 LRGB stars are selected for APOGEE and LAMOST, respectively.
However, the number of distant stars ($d > 8.5$\,kpc), although selected with loosed cuts, are very limited (only 293 and 134 for APOGEE and LAMOST).  
To include more distant stars in the training and testing samples, the member stars of globular clusters (GCs) are adopted.
The GC members are selected by the following cuts:
\\$\bullet$ spectral SNRs\,$> 20$;
\\$\bullet$ The position must within 15 half-light radius $r_h$  from the center of the GC;
\\$\bullet$ The proper motions must be $\left|\mu_{\rm{\alpha}}-\mu_{\rm{\alpha,GC}}\right| $ $\leqslant10 \sigma_{\mu_{\rm{\alpha,GC}}}$ and $\left|\mu_{\rm{\delta}}-\mu_{\rm{\delta,GC}}\right| $ $\leqslant10 \sigma_{\mu_{\rm{\delta,GC}}}$.

The above cuts are applied to 150 GCs with half-light radius adopted from \citet{2010arXiv1012.3224H} and proper motions measured by \citet{2019MNRAS.484.2832V}.
By further excluding few outliers in line-of-sight velocities, 2,383 members with $M_{K_{\rm s}} < -1.0$\,mag from 38 GCs are selected from the APO-LRGB sample.
The $K_{\rm s}$ band absolute magnitudes of those stars are derived by using the values of distance and extinction given in \citet{2010arXiv1012.3224H}.
In addition to the distant halo stars, 223 distant disk stars are selected with $d >$ 8.5\,kpc, $M_{K_{\rm s}} < -1.0$\,mag, spectral SNRs $> 50$, $|b| < 6^{\circ}$ and $e_{M_{K_{\rm s}}} <0.6$\,mag.
The above selections yield a sample of 10,145 LRGB unique stars (after removals of duplicate entries).
This sample is then divided into training sample with  7,101 stars and testing sample with 3,044 stars, respectively.

For the LM-LRGB sample, the number of GC members are no more than one hundred. 
To increase the number of both distant disk and halo stars in the training san testing samples, common LRGB stars between LAMOST and APOGEE are adopted.
A total of 2,359 LRGB stars are then selected with $M_{K_{\rm s}} < -1.0$\,mag, spectral SNRs $> 50$, and $6 < d < 20$\,kpc.
Amongst those stars, about one thousand distant halo stars with [Fe/H]\,$< -1.0$ and about 350 distant disk stars with $|b| < 10^{\circ}$ are included.
The $K_{\rm s}$ band absolute magnitudes of those stars are derived from the APOGEE spectra with the trained relations by the GBDT technique (see Section\,3.1).
In total, a sample of 9,063 unique LRGB stars are obtained for LAMOST.
Similar to APOGEE, the sample is divided into training sample with  6,484 stars and testing sample with 2,779 stars.
\end{appendix}

\end{document}